\documentclass[11pt,a4paper]{article}
\usepackage[a4paper,margin=1in]{geometry}
\usepackage{jheppub}
\usepackage[T1]{fontenc}
\usepackage[english]{babel}
\usepackage[usenames,dvipsnames]{xcolor}
\usepackage{comment}
\usepackage{float,graphicx,subfigure}
\usepackage{indentfirst,enumitem,multirow,appendix,caption,amsmath,amssymb,mathtools,braket,bm,bbm,nicefrac}

\title{Quantum detection of new physics in top-quark pair production at the LHC}

\author[a,b,c]{Fabio Maltoni,}
\author[d]{Claudio Severi,}
\author[c]{Simone Tentori,}
\author[d]{Eleni Vryonidou}

\affiliation[a]{Dipartimento di Fisica e Astronomia, Universit\`a di Bologna, via Irnerio 46, 40126 Bologna, Italy}
\affiliation[b]{INFN, Sezione di Bologna, Bologna, Italy}
\affiliation[c]{Center for Cosmology, Particle Physics and Phenomenology, Université Catholique de Louvain, Louvain-la-Neuve, Belgium}
\affiliation[d]{Department of Physics and Astronomy, University of Manchester, Oxford Road, Manchester M13~9PL, United Kingdom}

\emailAdd{fabio.maltoni@unibo.it}
\emailAdd{claudio.severi@manchester.ac.uk}
\emailAdd{simone.tentori@uclouvain.be}
\emailAdd{eleni.vryonidou@manchester.ac.uk}

\preprint{
\begin{flushright}
\end{flushright}
}

\abstract{The recent observation of entanglement between top and anti-top quarks at the LHC opens the way to interpreting collider data with quantum information tools. In this work we investigate the relevance of quantum observables in searches of new physics. To this aim, we study spin correlations of $t \bar t$ pairs originating from various intermediate resonances, and compare the discovery reach of quantum observables compared to classical ones. We find that they provide complementary information and, in several notable cases, also the additional leverage necessary to detect new
effects.}

\begin{document}
\maketitle

\clearpage

\section{Introduction}

Quantum Mechanics (QM), one of the most counter-intuitive and vanguard descriptions of fundamental phenomenoma ever conceived, is not only at the heart of our understanding of the Universe, of matter, and of its interactions, but has also gained a primary role in science and technology with a large range of applications to our everyday life going from computing, to information theory, to safe communications. 

While we currently have no motivation to think that QM would stop to describe phenomena at short distances, at least below the Planck scale, it is interesting to ponder to what extent fundamental quantum effects can be probed beyond the atomic scales ($10^{-10}$ m). Such a question has recently gained momentum after the observation of entanglement in the spin of top/anti-top quark pairs at the LHC \cite{ATLAS:2023fsd}, the highest energy accelerator experiment on earth, operating at the TeV ($10^{-19}$ m, $10^{-28}$ s) scale.

Numerous studies based on simulations \cite{2003.02280, 2102.11883, 2106.01377, 2110.10112, 2203.05582, 2204.11063, 2205.00542, 2208.11723, 2209.03969, 2209.13441, 2209.13990, 2209.14033, 2211.10513, 2302.00683, 2305.07075, 2306.17247} have elaborated further on these ideas, providing evidence that several more quantum effects may be visible in data collected (and to be collected) at the LHC, as well as  at future colliders. Apart from establishing entanglement, these measurements could also  potentially detect a violation of Bell inequalities. A variety of final states has been considered, most notably top-anti-top quark pairs, but also electroweak boson pairs, tau lepton pairs, and more. 

The thermally hot, dense, highly charged, and rapidly expanding environment produced by a high energy particle collision also yields an interesting setting to study the preservation of entanglement and the eventual decoherence.  
In addition, more exciting than just confirming expectations from QFT, quantum observables could provide additional leverage to search for beyond-the-Standard-Model (BSM) physics. Present collider data  generally indicate that new physics is either light and very weakly coupled to SM states, or lures beyond the TeV scale. 

In this work we explore a quite popular scenario at the LHC, where new physics is naturally connected to the top quark, and investigate the sensitivity of quantum observables to detect it in the production of top quark pairs.  The top quark is the heaviest known fundamental particle, and its large coupling to the Higgs boson places it in a unique spot for the study of SM and BSM physics. Moreover, with order million top quarks produced per inverse femtobarn of luminosity, experiments at the LHC are able to perform a plethora of measurements involving top quarks, ranging from very detailed properties of $t\bar t$ and single-top production, to the observations of more rare processes such as associated production of electroweak bosons and recently even the simultaneous production of four top quarks~\cite{2303.15061,2305.13439}. It is particularly exciting to realise that thanks to the existence of a such broad range of experimental results, some of which with high statistics, we are also presented with the unique opportunity to probe the top sector with quantum observables. 

Quantum effects in top/anti-top quark pairs are accessible through their spin. Top quarks are spin-$\nicefrac 1 2$ particles, and being mostly produced in pairs, they make an ideal two-qubit system at the TeV scale. Further, thanks to their large mass and therefore limited lifetime, strong interactions do not have time neither to decohere their quantum state nor to let them hadronise. Their quantum information is therefore predicted to be fully transmitted to the top decay products. 
Due to the properties of QCD and the experimental setup, at the LHC top quark pairs are produced with negligible individual spin polarisation but significant spin correlations. The existence of top-quark spin correlations in $t\bar t$ final state at the LHC has been established in 2012 \cite{1203.4081}. Spin correlations in $t \bar t$ were subsequently measured by the ATLAS and CMS Collaborations at a collision energy of $\sqrt{s} = 7 \, \text{TeV}$ \cite{ATLAS:2014aus}, $8 \, \text{TeV}$ \cite{CMS:2015cal, CMS:2016piu, ATLAS:2016bac}, and $13 \, \text{TeV}$ \cite{CMS:2019nrx, ATLAS:2019hau, ATLAS:2019zrq}, including several different top decay topologies. Further recent measurements at $13 \, \text{TeV}$ explored the dependence of spin correlations on kinematical properties of the top-quark pair, and, by measuring spin correlations near $t \bar t$ production threshold, established the presence of entanglement \cite{ATLAS:2023fsd}. Several further studies along the same lines are expected to appear in the near future.

Given the prospects of measuring quantum observables in top-quark final states with even higher statistics in the upcoming LHC runs \cite{CMS:2022cqm}, it is worth exploring whether these observables provide new or potentially enhanced sensitivity to new physics effects. In this work we set to study how new physics or unexpected SM effects affecting top-quark pair production could reveal themselves in quantum observables. Equipped with this information we determine how quantum observables complement classical kinematic observables, such as the invariant mass distribution of the top/anti-top pair, in the search for deviations from SM predictions.  

This paper is organized as follows. After a review of the results needed to describe and experimentally measure spin correlations in Section \ref{sec:setup}, we calculate analytically the $t\bar t$ spin correlation matrix for SM top-quark production in Section \ref{sec:sm} and in the presence of new resonant states in Section \ref{sec:particles}.  In Section \ref{sec:analysis} we perform a simulated analysis of a measurement of classical observables, related to the kinematics of the $t \bar t$ pair, and of quantum observables, the markers of entanglement commonly called $D$ as well as other similar quantities. We show that in interesting and realistic scenarios quantum effects offer significant advantage in detecting new physics or subtle SM effects. We provide our conclusions in Section \ref{sec:conclusions}.

\section{Top/anti-top quark pair spin correlations} \label{sec:setup}

Our description of fundamental interactions, as encapsulated in the SM, predicts that in a collider top/anti-top quark pairs are produced localised at very short distance in specific quantum states. Upon being produced, the top and the anti-top quarks have some time to fly apart  (with a well known distribution of relative speed) before decaying. Yet, in some areas of phase space, the top quarks remain ``connected" through their quantum wave function, i.e., they are entangled, and therefore exhibit spin correlation patterns that cannot be explained classically. 

A quantitative description of spin correlations of two spin-$\nicefrac 1 2$ particles needs nine degrees of freedom. Spin correlations are described by a $3 \times 3$ correlation matrix of real numbers,
\begin{equation}
    \mathcal C = \lbrace C_{ij} \rbrace \, _{i,j = 1,2,3}
\end{equation}
whose $ij$-entry represents the correlation between the $i$-th component of the top-quark spin and the $j$-th component of the anti-top-quark spin, $-1 \leq C_{ij} \leq 1$. 
Individual top-quarks may also be polarised, introducing six more degrees of freedom,
\begin{equation}
    \mathcal B_1 = \lbrace B_{1i} \rbrace \, _{i=1,2,3}, \quad  \mathcal B_2 = \lbrace B_{2j} \rbrace \, _{j=1,2,3}
\end{equation}
describing the average polarisation of the first and second particle along the $i$-th and $j$-th axes respectively. The spin density matrix is then given by: 
\begin{equation}
    \rho = \frac{1}{4} \big( \mathbf{1} \otimes \mathbf{1} + \mathcal B_{1} \cdot  \boldsymbol{\sigma} \otimes \mathbf{1} + \mathcal B_{2} \cdot  \mathbf{1} \otimes \boldsymbol{\sigma}  + \mathcal C \cdot \boldsymbol{\sigma} \otimes \boldsymbol{\sigma} \big), \label{rho}
\end{equation}
where $\boldsymbol{\sigma} = (\sigma_1, \sigma_2, \sigma_3 )$ are the Pauli matrices. Conservation of CP in $t \bar t$ production implies
\cite{Bernreuther:2015yna}:
\begin{equation}
    \mathcal C = \mathcal C^{\sf T}, \quad  \mathcal B_{1} = \mathcal B_{2} \label{cp}.
\end{equation}
We also note that separate conservation of $C$ and $P$ implies the stronger condition $\mathcal B_{1} = \mathcal B_{2} = 0$, which holds for the leading QCD production channels. 

\begin{figure}[H]
\centering
 \includegraphics[width=0.33\textwidth]{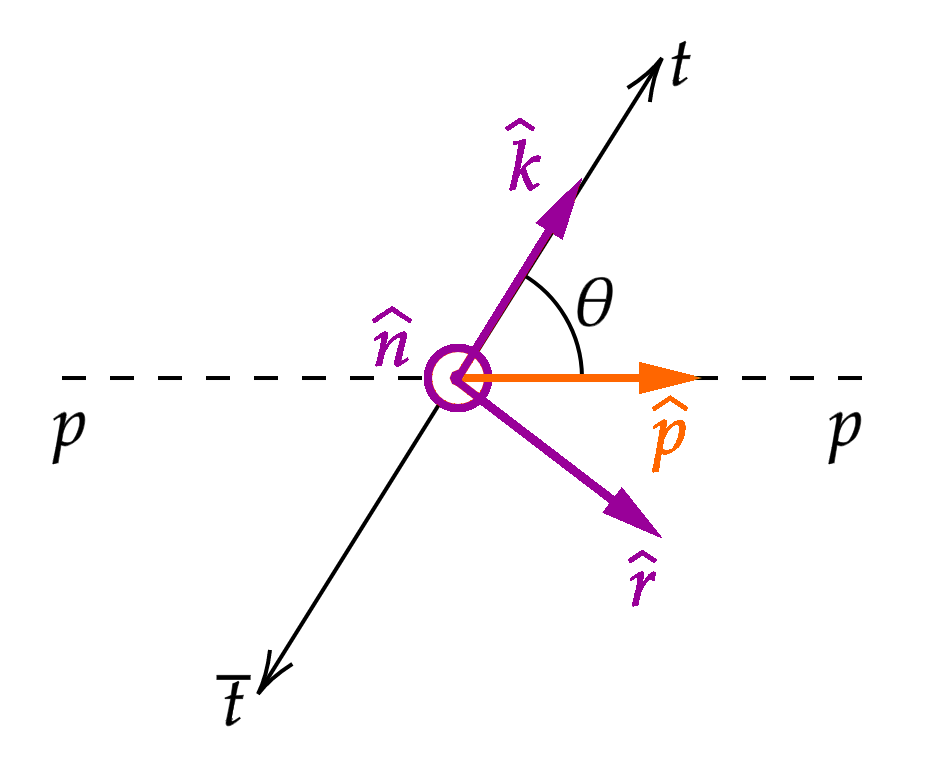}
  \caption{ Schematic representation of the production of a $t \bar t$ pair in its reference frame, with the helicity basis overlayed.}
 \label{fig:theta}
\end{figure}

To explicitly determine the coefficients, a basis must be chosen. In this work we will use the helicity basis $\lbrace \hat k, \hat r, \hat n \rbrace$, defined in the $t \bar t$ pair reference frame as
	\begin{equation}
		\hat k = \text{top direction}, \quad	\hat r = \frac{ \hat p - \hat k \, \cos \theta}{\sin \theta}, \quad \hat n = \frac{\hat p \times \hat k}{\sin \theta}	 \label{helbasis} \,,
	\end{equation}
	 where $\hat p$ is the beam axis and $\theta$ is the top scattering angle in the pair rest frame. We take $\theta$ to be from $0$ to $\pi/2$, see also Figure \ref{fig:theta}. 

The information on the quantum correlations of the $t\bar t$ spin state is fully contained in the matrix $\mathcal C$, and possibly relevant quantum observables, including entanglement, discord, steerability, concurrence, and Bell non-locality, can be computed from it \cite{2003.02280, 2102.11883, 2110.10112, 2203.05619, 2203.05582, 2205.00542, 2209.03969}. Among all, entanglement is the most readily accessible, and it is the one we shall focus on in the rest of this work. An application of the Peres-Horodecki criterion \cite{quant-ph/9604005, quant-ph/9605038}
shows that {\it sufficient} conditions for entanglement are:
\begin{align}
    - C_{kk} - C_{rr} - C_{nn} &> 1, \label{ent1} \\
    - C_{kk} + C_{rr} + C_{nn} &> 1, \\
    + C_{kk} - C_{rr} + C_{nn} &> 1, \\
    + C_{kk} + C_{rr} - C_{nn} &> 1.
    \label{ent4}
\end{align}
Any of the criteria \eqref{ent1}-\eqref{ent4} is sufficient for entanglement regardless of all other parameters in the density matrix. However, if the individual polarisations $\mathcal B$ or the off-diagonal elements of $\mathcal C$ are large, the conditions \eqref{ent1} only capture a certain region of parameter space where entanglement is present. This is not a problem in practice, since SM top-quark production in present colliders produces pairs with negligible individual polarisation and an almost diagonal correlation matrix. 

The four quantities
\begin{align}
    - C_{kk} - C_{rr} - C_{nn} &\equiv -3 \, D^{(1)}, \label{D1} \\
    - C_{kk} + C_{rr} + C_{nn} &\equiv -3 \, D^{(k)}, \\
    + C_{kk} - C_{rr} + C_{nn} &\equiv -3 \, D^{(r)}, \\
    + C_{kk} + C_{rr} - C_{nn} &\equiv -3 \, D^{(n)}.\label{D4}
\end{align}
are therefore the markers of spin entanglement most convenient to measure in a realistic collider scenario \footnote{We note that $D^{(1)}$ is usually called just $D$, while the triplet $D^{(k)}, D^{(r)}, D^{(n)}$ has been collectively denoted $D_3$ in \cite{2205.00542}.}. Entanglement is then signalled by any of the $D$'s satisfying the condition $D < -\nicefrac 1 3$.

The quantum state is pure when $\text{Tr} \, \rho^2 = 1$. Under the assumption of no net polarisation, when the quantum state is pure, it is also maximally entangled. In fact, taking the partial trace of $\rho$ over the top-quark or over the anti-top-quark subspaces yields:
\begin{equation}
    \rho_t = \rho_{\bar t} = \begin{pmatrix}
        \nicefrac 1 2 & 0 \\ 0 & \nicefrac 1 2
    \end{pmatrix}.
\end{equation}
A basis of pure, maximally entangled states for the $t \bar t$ spin described by \eqref{rho} is given by the four Bell states:
\begin{align}
    \ket{\Phi^\pm} &= \frac{1}{\sqrt 2}\big( \ket{\uparrow \uparrow} \pm \ket{\downarrow \downarrow} \big) \label{bell1}, \\
    \ket{\Psi^\pm} &= \frac{1}{\sqrt 2}\big( \ket{\uparrow \downarrow} \pm \ket{\downarrow \uparrow} \big) \label{bell2}.
\end{align}
We note that the limiting cases
\begin{equation}
    \mathcal C^{(\text{singlet})} = 
    \begin{pmatrix}
     -\eta & 0 & 0 \\
     0 & -\eta & 0 \\
     0 & 0 & -\eta
    \end{pmatrix}, \qquad  \mathcal C^{(\text{triplet})} = 
    \begin{pmatrix}
     \eta & 0 & 0 \\
     0 & \eta & 0 \\
     0 & 0 & -\eta
    \end{pmatrix}, \qquad 0 < \eta \leq 1 \, 
\end{equation}
and cyclic permutations of the diagonal of $\mathcal C^{(\text{triplet})}$, correspond to the singlet (left) and triplet (right) Werner states. 
The case of a quantum state described by $\mathcal C^{(\text{singlet})}$ is rather special, as a spin measurement on such a state yields anticorrelated results, with anticorrelation $\eta$, along {\it all} possible measuring axes (not just those in the chosen basis). 

We also note that for Werner states one of the entanglement markers \eqref{D1} - \eqref{D4} reduces to:
\begin{equation}
    D = -\eta,
\end{equation}
specifically $D^{(1)} = -\eta$ for the singlet, and $D^{(i)} = -\eta$ for the three triplets.
For Werner states, it is known that
\begin{align}
    \eta > \frac{1}{3} \ &\implies \ \text{entanglement}, \\
    \eta > \frac{1}{\sqrt{2}} \ &\implies \ \text{Bell inequality violation}, \\
    \eta = 1 \ &\implies \ \text{pure state}. 
\end{align}
In the limit $\eta \to 1$ the Werner states reduce to pure singlet $\ket{\Psi^-}$ and pure triplet $\ket{\Psi^+}$, $\ket{\Phi^-}$, $\ket{\Phi^+}$ Bell states. 

    The spin state of particles decaying electroweakly is transferred into the direction of flight of their decay products. Therefore, the spin state of $t \bar t$ pairs survives their decay, and leaves an imprint in the direction of flight of top-quark decay products. Spin correlations of top quarks are accessible experimentally from the correlations of angles between their daughters, i.e.\@ leptons or jets. To recover the information about the spin, the relevant angles have to be evaluated in the top-quark rest frame to be reached from the $t \bar t$ zero momentum frame with a rotation-free boost. Reconstructing the two top quarks rest frames therefore requires knowledge of all six of their decay products, {\it i.e.}, two $b$--jets, and, depending on the selected decay channel, up to four light jets, charged leptons, or neutrinos. We note that the top rest frames can always be reconstructed from experimentally accessible information, even when two neutrinos are present in the final state, see for instance the techniques used by the ATLAS and CMS Collaborations in the recent $t \bar t$ spin analyses \cite{ATLAS:2023fsd,CMS:2019nrx}.

    Assuming the top quark decays into particle $a$ (plus other particles) and the anti-top decays into particle $b$, the differential cross section for $t \bar t$ production plus the decays $t \to a+X$ and $\bar t \to b+X$ is given at LO by
    \cite{Bernreuther:2015yna}:
	\begin{equation}
	    \frac{1}{\sigma} \frac{d \sigma}{d (\cos \theta_{ai} \cos \theta_{bj})} = -\frac{1 + C_{ij} \, \alpha_a \, \alpha_b \, \cos \theta_{ai} \cos  \theta_{bj}}{2} \, \log \big|\cos \theta_{ai} \cos  \theta_{bj} \, \big| \label{dcoscos}
	\end{equation}
	 (no sum over $i$ or $j$), where $\theta_{ai}$ is the angle between the momentum of $a$ and the $i$-th axis in the top rest frame, and $\theta_{bj}$ is the angle between the momentum of $b$ and the $j$-th axis in the anti-top rest frame. The parameters $\alpha_a$ and $\alpha_b$ provide a measure of the  {\it spin analyzing power} of particles $a$ and $b$, that is, they parameterize how much their direction of emission is correlated to the original top/anti-top-quark spin. The spin analyzing power of top decay products in the SM is given in Table \ref{tab:spinpower}.

    \begin{table}[h]
	\centering
	\begin{tabular}{|c c|} 
 		\hline
 		Particle & $\alpha$ \\ [0.5ex] 
 		\hline \hline
		$b$ & $-0.3925(6)$ \\
		$W^+$ & $\ \ 0.3925(6)$ \\
 		$\ell^+$ (from a $W^+$) & $\, 0.999(1)$  \\ 
 		$\bar d, \, \bar s$ (from a $W^+$) & $\ \ 0.9664(7)$ \\
		$u, \, c$ (from a $W^+$) & $-0.3167(6)$ \\
 		\hline
	\end{tabular}
	\caption{Spin analyzing power of top decay products in the SM at NLO accuracy \cite{Czarnecki:1990pe}, \cite{Brandenburg:2002xr}. Values for antiparticles differ by a sign.}
	\label{tab:spinpower}
\end{table}
 
 In the SM, light charged leptons have $\alpha \simeq 1$, making the dileptonic channel very promising for top spin correlations studies. As shown in \cite{2210.09330}, heavy (with respect to the top mass) new physics is unlikely to fundamentally alter this picture.

    The integration of \eqref{dcoscos} gives an explicit relation for the entries of $\mathcal C$ in terms of the average value of $\cos \theta_{ai} \cos \theta_{bj}$:
	\begin{equation}
		C_{ij} = \frac{9}{\alpha_a \alpha_b} \> \texttt{Avg} \left[ \cos \theta_{ai} \cos \theta_{bj} \right]. \label{dcos_integrated}
	\end{equation}
    In an experiment the average is taken over real events, inclusively or differentially in suitable kinematical variables. We note that, apart from extra radiation, there are only two degrees of freedom in the $t \bar t$ production kinematics: the top pair invariant mass $m_{t \bar t}$ and the scattering angle $\theta$ with respect to the beam. 
    
    Analytically, the average in \eqref{dcos_integrated} is taken in phase space, weighted by the matrix element squared. For instance, the entries of $\mathcal C$ as a function of $m_{t \bar t}$ and $\theta$ for the partonic process $i_1 \, i_2 \to t \, \bar t$ is obtained as:
	\begin{equation}
    C_{ij}^{[i_1 i_2]}(m_{t \bar t}, \theta) =  \frac{\nicefrac{9}{\alpha_a \alpha_b} \int \cos \theta_{ai} \cos \theta_{bj} \, |\mathcal M_{i_1 \, i_2 \to t \, \bar t \to a \, b \, X}|^2 \, d\pi} {\int |\mathcal M_{i_1 \, i_2 \to t \, \bar t \to a \, b \, X}|^2 \, d\pi},
    \label{xsec_ratio}
	\end{equation}
     where the integration $d \pi$ is on the final state phase space region at constant $m_{t \bar t}$ and $\theta$. Of course, if $\mathcal C^{[i_1 i_2]}$ is needed as a function of other variables, or inclusively, the integration $d\pi$ has to be adapted accordingly. Therefore, following the notation of \cite{2203.05619}, it is natural to parameterize the spin correlation matrix as:
\begin{equation}
\mathcal C = \frac{\widetilde{\mathcal C}}{A}. \label{cij}
\end{equation}
The matrix $\widetilde{\mathcal C}$ and the common normalisation $A$ have units of cross-sections, and are given by the numerator and denominator of \eqref{xsec_ratio}. The decomposition \eqref{cij} is general, and irrespective of e.g.\@ the underlying top-quark production channels or phase-space cuts. 

If several non-interfering $t \bar t$ production channels $1, 2, \cdots$ are available for the same partonic process (e.g.\@ for $u \bar u \to t \bar t$ one may consider strong $u \bar u \to g^* \to t \bar t$ and electroweak $u \bar u \to \gamma^* /Z^* \to t \bar t$ production), spin correlations are calculated as:
\begin{align}
   \mathcal C^{[i_1 i_2]} &= \frac{\sum_{k} \widetilde{\mathcal C}^{[i_1 i_2, \, k]}}{\sum_l A^{[i_1 i_2, \, l]}}
    ={\sum_k w_k \, \mathcal C^{[i_1 i_2, \, k]}} \quad {\rm with} \quad
w_k \equiv \frac{A^{[i_1 i_2, \, k]}}{\sum_l A^{[i_1 i_2, \, l]}}, \label{spinobs}
\end{align}
{\it i.e.}, spin correlations are obtained as the weighted average of the spin correlations $\mathcal C^{[i_1 i_2, \, k]}$ stemming from each channel $k$, weighted by $w_k$, proportional to the corresponding partonic matrix element squared
$A^{[i_1 i_2,\, k]}$.

 On top of this, in proton-proton collisions, several partonic processes enter in top pair production, each one potentially receiving multiple internal contributions as in \eqref{spinobs}. The $A$ and $\widetilde{\mathcal C}$ of the complete process $p p \to t \bar t$ are then obtained by rescaling those of partonic processes by the corresponding luminosity:
 \begin{align}
     A^{[pp]}(m_{t \bar t}, \theta) &= \sum_{i_1,i_2} L_{i_1,i_2}(m_{t \bar t}, \sqrt{s}) \> A^{[i_1 i_2]}(m_{t \bar t}, \theta), \\
     \widetilde{\mathcal C}^{[pp]}(m_{t \bar t}, \theta) &= \sum_{i_1,i_2} L_{i_1,i_2}(m_{t \bar t}, \sqrt{s}) \> \widetilde{\mathcal C}^{[i_1 i_2]}(m_{t \bar t}, \theta),
 \end{align}
where $L_{i_1,i_2}(m_{t \bar t},\sqrt{s})$ parameterizes the probability to obtain the $i_1,i_2$ initial state at partonic energy $m_{t \bar t}$ in a proton-proton collision at total energy $\sqrt{s}$, see \cite{Bernreuther:1997gs} for the precise definition of $L$ in terms of PDFs.

In the following sections we examine $t \bar t$ spin correlations stemming from QCD, and $s$-channel exchange of a photon and $Z$ boson, that allow us to study vector ($\gamma^\mu$) and axial vector ($\gamma^\mu \gamma^5$) interactions of the top quark. We also consider some simple BSM scenarios that allow us to analyse top-quark production through scalar ($\mathbbm{1}$) and pseudoscalar ($\gamma^5$) interactions.

\section{Spin correlations in the Standard Model} \label{sec:sm}

SM top/anti-top quark pair production, neglecting $b$-quark initiated processes,  happens at tree level via two non-interfering channels, one mediated by QCD at order $\alpha_{s}^2$, and one mediated by electroweak interactions, at order $\alpha^2$. 

\subsection{QCD top production} \label{sec:sm_qcd}

Spin correlations in $t \bar t$ pairs produced by QCD have been known for a long time, and are now available numerically at NNLO \cite{1901.05407, Czakon:2020qbd}. Analytical evaluation of the spin correlation matrix for the short distance part is available at LO. Nevertheless, it is known \cite{1508.05271} that in the SM higher order corrections to spin observables are small, and the LO result tends to capture the essence of the physical picture. For completeness we report here the spin correlation coefficients at LO. For gluon fusion $g g \to t \bar t$, the spin correlation coefficients are given by:
\begin{align}
A^{[gg, \, \text{\tiny QCD}]} &= F^{[gg, \, \text{\tiny QCD}]} \left[ -\beta^4 c_{\theta }^4 +2 \left(\beta^2-1\right) \beta^2 c_{\theta }^2-2 \beta^4+2 \beta^2+1 \right], \label{smqcdgg1} \\
\widetilde{C}^{[gg, \, \text{\tiny QCD}]}_{kk} &=  F^{[gg, \, \text{\tiny QCD}]} \left[ \beta^2 c_{\theta }^2 \left(\left(\beta^2-2\right) c_{\theta }^2-2 \beta^2+2\right)+2 \beta^4-1  \right)], \\
\widetilde{C}^{[gg, \, \text{\tiny QCD}]}_{kr} &= F^{[gg, \, \text{\tiny QCD}]} \, 2 \beta^2 \sqrt{1-\beta^2} \, c_{\theta } s_{\theta }^3 , \\
\widetilde{C}^{[gg, \, \text{\tiny QCD}]}_{rr} &= F^{[gg, \, \text{\tiny QCD}]} \left[ -\beta^4 c_{\theta }^2 \left(c_{\theta }^2-2\right)+2 \beta^2 s_{\theta }^4-2 \beta^4+2 \beta^2-1   \right], \\
\widetilde{C}^{[gg, \, \text{\tiny QCD}]}_{nn} &= F^{[gg, \, \text{\tiny QCD}]} \left[- \beta^4 c_{\theta }^2 \left(c_{\theta }^2-2\right)-2 \beta^4+2 \beta^2-1    \right], \label{smqcdggn}
\end{align}
while for $q \bar q \to t \bar t$ they are given by:
\begin{align}
A^{[q\bar q, \, \text{\tiny QCD}]} &= F^{[q\bar q, \, \text{\tiny QCD}]}\left( \beta^2 c_{\theta }^2-\beta^2+2  \right), \label{smqcdqq1} \\
\widetilde{C}^{[q\bar q, \, \text{\tiny QCD}]}_{kk} &= F^{[q\bar q, \, \text{\tiny QCD}]} \left[ \beta^2-\left(\beta^2-2\right) c_{\theta }^2  \right], \\
\widetilde{C}^{[q\bar q, \, \text{\tiny QCD}]}_{kr} &= F^{[q\bar q, \, \text{\tiny QCD}]} \, 2 \sqrt{1-\beta^2} c_{\theta } s_{\theta } , \\
\widetilde{C}^{[q\bar q, \, \text{\tiny QCD}]}_{rr} &= F^{[q\bar q, \, \text{\tiny QCD}]} \, s_{\theta }^2 (2 -\beta^2), \\
\widetilde{C}^{[q\bar q, \, \text{\tiny QCD}]}_{nn} &=  - F^{[q\bar q, \, \text{\tiny QCD}]}  \, \beta^2 s_{\theta }^2\,.\label{smqcdqqn} 
\end{align}
We use the shorthand notation $\cos \theta \equiv c_\theta$ and $\sin \theta \equiv s_\theta$ for the cosine and sine of the top scattering angle in the $t \bar t$ rest frame, and for convenience we have used as the second kinematical variable the top velocity $\beta$, given by:
    \begin{equation}
        \beta^2 = 1-\frac{4 m_t^2}{m_{t \bar t}^2}.
    \end{equation}

Throughout this work, we collect the common factors between $A$ and $\widetilde{\mathcal C}$ in symbols denoted with $F$. The explicit expressions for $F^{[gg, \, \text{\tiny QCD}]}$ and $F^{[q\bar q, \, \text{\tiny QCD}]}$ are:
\begin{align}
    F^{[gg, \, \text{\tiny QCD}]} &= \frac{16 g_s^4 \left(9 \beta^2 c_{\theta }^2+7\right)}{3 \left(\beta^2 c_{\theta }^2-1\right)^2}, \label{fqcdgg}\\
    F^{[q\bar q, \, \text{\tiny QCD}]} &= 32 g_s^4 \label{fqcdqq}.
\end{align}

The results in \eqref{smqcdgg1} - \eqref{smqcdqqn} agree with previous work~\cite{2003.02280, 2203.05619}, and have also been confirmed numerically with a simulation based on the Monte Carlo generator {\tt MadGraph5\_aMC@NLO} \cite{1405.0301} and the code developed in \cite{2110.10112}. Our method of analytical extraction of $\mathcal C$ from helicity amplitudes is described in Appendix \ref{app:spin_calc}. 

\subsection{Electroweak top-quark production} \label{sec:sm_ew}

\begin{figure}[H]
    \centering
    \includegraphics[width=.65\textwidth]{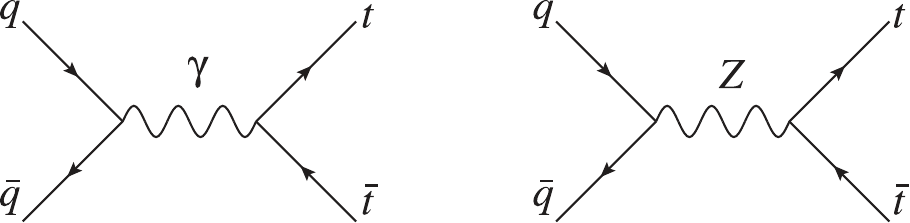}
    \caption{Feynman diagrams corresponding to electroweak $t \bar t$ production. }
    \label{fig:ewdiags}
\end{figure}

Top-quark pair production can also proceed through weak interactions.  Electroweak $t \bar t$ production is described at LO by the diagrams in Figure \ref{fig:ewdiags}. To maintain generality, we describe the initial state with its charge $Q_i$ and isospin $T_{3i}$, so that our results can be applied to any electroweak top-quark production process. We write the electroweak vertices between two fermions and a vector as \footnote{Since the notation may be ambiguous due to the presence of Dirac $\gamma$ matrices, we indicate the photon with the letter $\mathcal A$.}
\begin{equation}
    \left( f \, \bar f \, \mathcal A^\mu \right) = i e Q_f \, \gamma^\mu, \qquad \left( f \, \bar f \, Z^\mu \right) = \frac{i e}{c_W s_W}  \, \gamma^\mu \, \left( g_{\text{V}f} \, \mathbbm{1} - g_{\text{A}f} \, \gamma^5 \right) \,,
\end{equation}
where $g_\text{V} = T_3/2 - Q \, s_W^2$ and $g_\text{A} = T_3/2$, with $T_3 = +\nicefrac 1 2$ for up-type quarks, and $T_3 = -\nicefrac 1 2$ for down-type quarks.

To better highlight the structure of spin correlations, we split the electroweak matrix element squared as follows:
\begin{equation}
    A^{[q\bar q, \, \text{\tiny EW}]} = A^{[q\bar q, \, \text{\tiny EW}, \, 0]} + A^{[q\bar q, \, \text{\tiny EW}, \, 1]} + A^{[q\bar q, \, \text{\tiny EW}, \, 2]},
\end{equation}
where the various terms of the squared amplitude contain, respectively, 
\begin{enumerate}
    \item[0.] All terms with no insertion of $\gamma^5$ in the $t \bar t$ fermion line, i.e.\@ terms proportional to $Q_t^2$, $g_{\text{V}t}^2$, and $Q_t \, g_{\text{V} t}$, are collected in $A^{[q\bar q, \, \text{\tiny EW}, \, 0]}$.
    \item[1.] All terms with a {\it single} insertion of $\gamma^5$ in the top-quark fermion line, i.e.\@ terms proportional to $g_{\text{A}t} Q_{t}$ and $g_{\text{A}t} g_{\text{V}t}$, contribute to $A^{[q\bar q, \, \text{\tiny EW}, \, 1]}$.
    \item[2.] All terms with a {\it double} insertion of $\gamma^5$ in the top-quark fermion line, i.e.\@ terms with $g_{\text{A}t}^2$, enter $A^{[q\bar q, \, \text{\tiny EW}, \, 2]}$.
\end{enumerate}

The $A^{[q\bar q, \, \text{\tiny EW}, \, 0]}$ channel yields the same spin correlations as QCD \eqref{smqcdqq1}-\eqref{smqcdqqn}, since the helicity structure is the same. The only difference is the normalisation factor $F$, which here is given by the expression in \eqref{few0}.

The $A^{[q\bar q, \, \text{\tiny EW}, \, 1]}$ term produces:
\begin{align}
A^{[q\bar q, \, \text{\tiny EW}, \, 1]} &= 2 \, F^{[q\bar q, \, \text{\tiny EW}, \, 1]} \, c_\theta, \label{smaqq1} \\
\widetilde{C}_{kk}^{[q\bar q, \, \text{\tiny EW}, \, 1]} &= 2 \, F^{[q\bar q, \, \text{\tiny EW}, \, 1]} \, c_\theta, \\
\widetilde{C}_{kr}^{[q\bar q, \, \text{\tiny EW}, \, 1]} &= F^{[q\bar q, \, \text{\tiny EW}, \, 1]} \, \sqrt{1-\beta^2} \, s_\theta, \\
\widetilde{C}_{rr}^{[q\bar q, \, \text{\tiny EW}, \, 1]} &= 0, \\
\widetilde{C}_{nn}^{[q\bar q, \, \text{\tiny EW}, \, 1]} &= 0, \label{smaqqn}
\end{align}

while the $A^{[q\bar q, \, \text{\tiny EW}, \, 2]}$ term gives:

\begin{align}
A^{[q\bar q, \, \text{\tiny EW}, \, 2]} &= F^{[q\bar q, \, \text{\tiny EW}, \, 2]} \, \left(1 + c_\theta^2   \right), \label{smavqq1} \\
\widetilde{C}_{kk}^{[q\bar q, \, \text{\tiny EW}, \, 2]} &= F^{[q\bar q, \, \text{\tiny EW}, \, 2]} \, \left( 1 + c_\theta^2   \right), \\
\widetilde{C}_{kr}^{[q\bar q, \, \text{\tiny EW}, \, 2]} &= 0, \\
\widetilde{C}_{rr}^{[q\bar q, \, \text{\tiny EW}, \, 2]} &= - F^{[q\bar q, \, \text{\tiny EW}, \, 2]} \, s_\theta^2, \\
\widetilde{C}_{nn}^{[q\bar q, \, \text{\tiny EW}, \, 2]} &= F^{[q\bar q, \, \text{\tiny EW}, \, 2]} \, s_\theta^2\,. \label{smavqqn}
\end{align}

The common factors for the ``$0$'', ``$1$'', and ``$2$'' channels of SM electroweak top-quark production are:
\begin{align}
    F^{[q\bar q, \, \text{\tiny EW}, \, 0]} &= 144 e^4 \Big( Q_t^2 Q_i^2 + 
    2 \, \text{Re} \, \frac{4 Q_t Q_i g_{\text{V} t} g_{\text{V} i} m_t^2}{c_W^2 s_W^2 \Pi_Z}+ \frac{16 g_{\text{V} t}^2 \left(g_{\text{A} i}^2+g_{\text{V} i}^2\right) m_t^4}{c_W^4 s_W^4 |\Pi_Z|^2}\Big), \label{few0} \\
   F^{[q\bar q, \, \text{\tiny EW}, \, 1]} &= 576 e^4 g_{\text{A} t} g_{\text{A} i} m_t^2 \beta \Big(\frac{16 g_{\text{V} t} g_{\text{V} i} m_t^2}{c_W^4 s_W^4 |\Pi_Z|^2} + 2 \, \text{Re} \, \frac{Q_t Q_i}{c_W^2 s_W^2 \Pi_Z} \Big), \label{few1} \\
    F^{[q\bar q, \, \text{\tiny EW}, \, 2]} &= \frac{2304 e^4 m_t^4 \beta^2 g_{\text{A} t}^2 \left(g_{\text{A} i}^2+g_{\text{V} i}^2\right)}{c_W^4 s_W^4 |\Pi_Z|^2} \label{few2}.   
\end{align}

We have denoted $\Pi_Z = 4 m_t^2+m_Z^2 \left(\beta^2-1\right)$.\footnote{The $Z$ boson propagator is $1 / (s - m_Z^2) = (1-\beta^2)/\Pi_Z$.} Finite width effects may be implemented by replacing $m_Z^2 \to m_Z^2 - i m_Z \Gamma_Z$ in $\Pi_Z$.  \smallskip

Whilst we have focused on the impact of EW interactions within the SM, the results shown in this section can be straightforwardly modified to account for new resonances coupling to the top quark with a vector or axial vector coupling. In the case of new narrow resonances, the only difference would be in the normalisation factors, that would be adjusted to account for the different masses and couplings.

\section{Spin correlations with new intermediate states} \label{sec:particles}

We now consider explicit new physics models, given by the extension of the SM with new particles light enough to be resonantly produced in collider experiments.

\subsection{Scalar/pseudoscalar resonances} \label{sec:scalar}

One interesting scenario to consider is the introduction of a spin-0 state $\phi$ that couples to SM tops with a scalar and pseudoscalar interaction, in a simplified model similar to the one studied in \cite{Frederix:2007gi, Bernreuther:2017yhg,Coloretti:2023yyq}:
\begin{equation}
    \mathcal L  = \mathcal L_{\text{SM}} - \frac{1}{2} \phi (\partial^2 + M_\phi^2) \phi + c_y \, \frac{y_t}{\sqrt{2}} \, \phi \, \overline t \, (\cos \alpha + i \gamma^5 \sin \alpha) \, t. \label{httlagrangian}
\end{equation}
In this simplified case there are only three parameters, the heavy scalar mass $M_\phi$, the coupling $c_y$ (normalised as a rescaling of the SM top-quark Yukawa coupling), and the angle $\alpha$, which produces a scalar particle for $\alpha = 0$ and a pseudoscalar particle for $\alpha = \pi/2$.

\begin{figure}[t]
\centering\includegraphics[width=.45\textwidth]{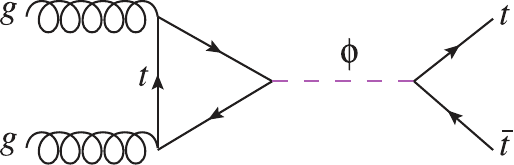}
    \caption{Feynman diagram corresponding to $t \bar t$ production mediated by a scalar $\phi$.}
    \label{fig:htt}
\end{figure}

At LO in $c_y$ top/anti-top quark pair production mediated by $\phi$ is given by the diagram in Figure \ref{fig:htt}.

The heavy scalar, like the SM Higgs, only couples to top quarks at tree-level, so that for $M_\phi > 2 m_t$ the largely dominant decay channel is the on-shell $t \bar t$ pair. Explicitly, the decay width $\Gamma_{\phi \to t \bar t }$ is given at LO by:
\begin{equation}
\Gamma_{\phi\rightarrow t\bar t}=\frac{3 \, c_y^2 \, y_t^2 \, M_\phi}{16\pi} \sqrt{1 - \frac{4m_t^2}{M_\phi^2}} \left(1 - \frac{4m_t^2}{M_\phi^2} \cos^2 \alpha\right). \label{htt_width}
\end{equation}
We note $\phi$ may also decay at one loop to a gluon pair, which is dominant for light scalars, but this channel is subdominant with respect to the tree-level process $\phi \to t \bar t$ when $M_\phi > 2 m_t$. 

The spin correlation coefficients for $t \bar t$ production mediated by $\phi$, that is for the square of the diagram in Figure \ref{fig:htt}, are given by:
\begin{align}
A^{[gg, \, \phi]} &= F^{[gg, \, \phi]} \left( s_\alpha^2 + \beta^2 c_\alpha^2 \right), \label{hsq1} \\
\widetilde{C}^{[gg, \, \phi]}_{kk} &= F^{[gg, \, \phi]} \left( - s_\alpha^2 - \beta^2 c_\alpha^2 \right), \\
\widetilde{C}^{[gg, \, \phi]}_{kr} &= 0, \\
\widetilde{C}^{[gg, \, \phi]}_{rr} &= F^{[gg, \, \phi]} \left( - s_\alpha^2 + \beta^2 c_\alpha^2 \right), \\
\widetilde{C}^{[gg, \, \phi]}_{nn} &= F^{[gg, \, \phi]} \left( - s_\alpha^2 + \beta^2 c_\alpha^2 \right). \label{hsqn}
\end{align}
We used the shorthand notation $c_\alpha \equiv \cos \alpha$ and $s_\alpha \equiv \sin \alpha $. Note that here $\alpha$ is a parameter, not the scattering angle $\theta$ of previous sections.  

In addition to \eqref{hsq1}--\eqref{hsqn}, for $\alpha \neq 0$ and $\alpha \neq \pi/2$ we also have the CP-violating correlations:
\begin{equation}
    \widetilde{C}^{[gg, \, \phi]}_{rn} = - \widetilde{C}^{[gg, \, \phi]}_{nr} = 2 F^{[gg, \, \phi]} \, \beta s_\alpha c_\alpha. 
\end{equation}

The normalisation factor $F^{[\phi]}$ is given by:
\begin{equation}
    F^{[gg, \, \phi]} = \frac{3 m_t^4 c_y^4 y_t^4 g_s^4 (1-\beta^2)}{8 \pi^4 |\Pi_\phi|^2} \ \text{Re} \left(  c_\alpha^2 \big(\beta^2 \log^2 \frac{\beta-1}{\beta+1} - 4 \big)^2 + s_\alpha^2 \log^4 \frac{\beta-1}{\beta+1} \right). \label{fh}
\end{equation}
We have denoted $\Pi_\phi = 4 m_t^2+m^2 \left(\beta^2-1\right)$. Similarly to the case of the $Z$ boson, finite width effects can be accounted for by replacing $m^2 \to m^2 - i m \Gamma$ in $\Pi_\phi$.

If the particle is a scalar, the $t \bar t$ quantum state is always a pure triplet, while if the particle is a pseudoscalar the $t \bar t$ quantum state is a pure singlet:
\begin{equation}
    \mathcal C^{[gg, \, \phi]} \big|_{\alpha = 0} = \begin{pmatrix} -1 & 0 & 0 \\ 0 & 1 & 0 \\ 0 & 0 & 1 \end{pmatrix}, \qquad \mathcal C^{[gg, \, \phi]} \big|_{\alpha = \pi/2} = \begin{pmatrix} -1 & 0 & 0 \\ 0 & -1 & 0 \\ 0 & 0 & -1 \end{pmatrix}. \label{H}
\end{equation}

In addition to the pure $\phi$ contribution, QCD and $\phi$ production can interfere in the $gg$ channel, and one might wonder about the quantum state of the top pair in this case. In fact, a calculation is not needed: since the Bell states \eqref{bell1}-\eqref{bell2} are orthogonal, when the $gg \to \phi \to t \bar t$ amplitude reduces to a pure Bell state it acts as a projector. Top-quark production via the exchange of a $\phi$, therefore, is described by the spin state \eqref{H} at both the interference and squared level. By the same argument, since the quantum states reached for $\alpha = 0$ and $\alpha = \pi/2$ are orthogonal, the scalar--pseudoscalar interference vanishes identically. One can easily check that the corresponding QFT amplitudes $gg \to \phi_{\alpha = 0} \to t \bar t$ and $gg \to \phi_{\alpha = \pi/2} \to t \bar t$ do not interfere for on-shell tops.


\subsubsection{Effective description of $t \bar t$ bound states} \label{sec:toponium}

\begin{figure}[H]
    \centering
    \includegraphics[width=0.6\textwidth]{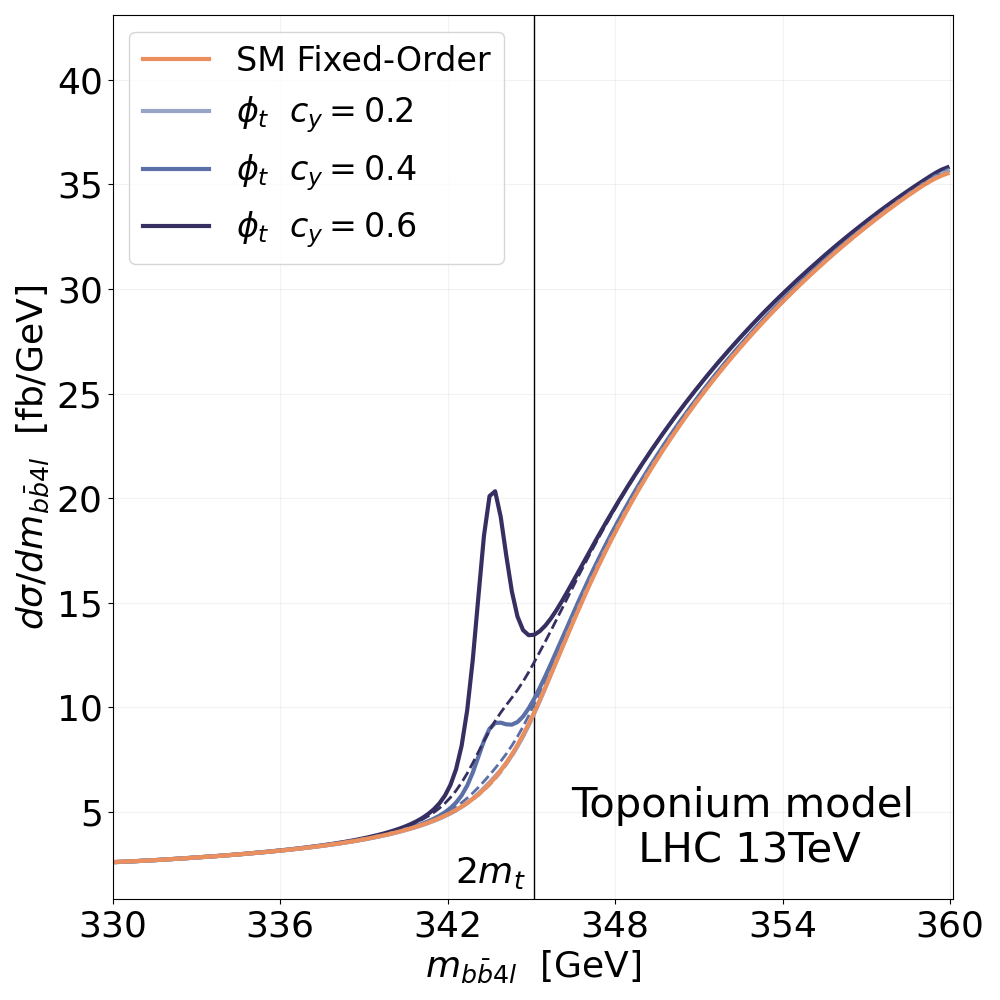}
    \caption{Differential cross section as a function of the invariant mass of the final state $m_{b\bar b 4\ell}$ for the process $p p \to b \, \bar b \, \ell^+ \, \ell^- \, \nu \, \bar \nu$. Orange: SM prediction, including non-resonant effects. Blue: pseudoscalar $\phi$ with $M_\phi = 343.5 \, \text{GeV}$. Solid line: $\Gamma_\phi = 1 \, \text{GeV}$, dashed line: $\Gamma_\phi = 2.5 \, \text{GeV}$. The coupling $c_y$ is as described in the legend. }
    \label{fig:line_threshold}
\end{figure}

\begin{figure}[H]
    \centering
    \includegraphics[width=0.6\textwidth]{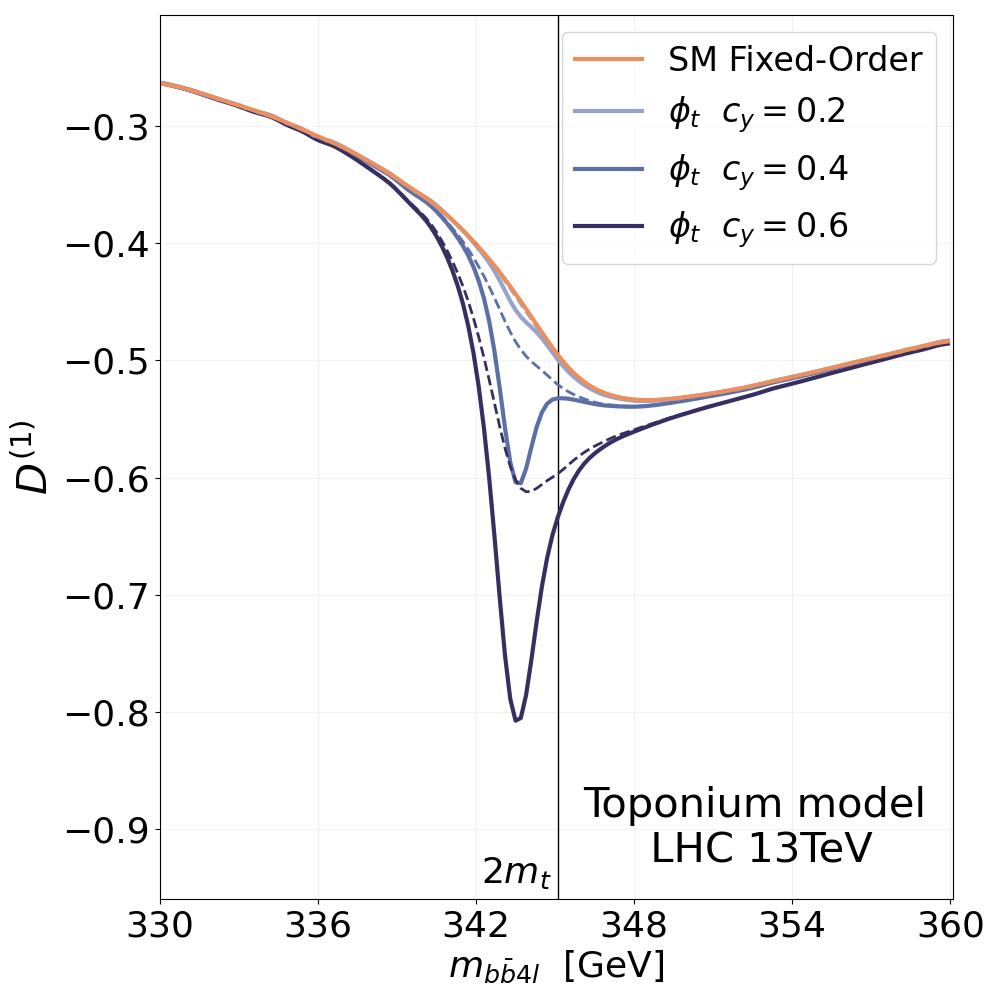}
    \caption{Same as Fig.~\ref{fig:line_threshold}, but considering the effect on $D^{(1)}$.}
    \label{fig:line_threshold_D}
\end{figure}

The model described in this Section is a typical BSM simplified model. However, it can also be taken as a simple effective description for the production of a $t \bar t$ bound state near threshold. Several studies \cite{0804.1014,0812.0919} have suggested that such a pseudo-bound state ({\it "toponium"}) leads to an enhancement of the cross section in the color singlet $g g \to t \bar t$ channel at threshold resulting in a structure that resembles a resonance peak.  The Lagrangian in Eq.~\eqref{httlagrangian} can then be used as a first rough model of such an enhancement as also suggested in Ref.~\cite{2102.11281}. We also note that recently, toponium effects  received further attention after the publication of the ATLAS entanglement measurement \cite{ATLAS:2023fsd}, which exhibits an interesting negative excess with respect to all NLO+PS predictions of $D^{(1)}$ near $t \bar t$ threshold.  Even though the resolution in invariant mass is not sufficient to draw any conclusion,  such an enhancement could be  consistent with the enhancement  predicted by QCD.  The Lagrangian in Eq.~\eqref{httlagrangian} allows to provide a modelization of such a pseudo-resonance, choosing parameters so to resemble the QCD predictions, {\it i.e.}, 
\begin{equation}
    M_\phi = 343.5 \, \text{GeV}, \quad \alpha = \pi/2.
\end{equation}
When taking \eqref{httlagrangian} as an effective description of toponium, the coupling $c_y$ and the width $\Gamma$ may be tuned independently to give a toponium cross-section consistent with theoretical QCD predictions. As an example, we show the invariant mass distribution $m_{b \bar b 4 \ell}$ and the value of $D^{(1)}$ resulting from the process:
\begin{equation}
    p p \to b \, \bar b \, \ell^+ \, \ell^- \, \nu \, \bar \nu
\end{equation}
in the SM, and in the presence of a toponium-like $\phi$ in Figures \ref{fig:line_threshold} and \ref{fig:line_threshold_D}.

We extract the SM background with {\tt MadGraph5\_aMC@NLO} at LO in QCD and the EW couplings, with the {\tt NNPDF4.0} parton distribution functions \cite{NNPDF:2021njg}, and factorization and renormalization scales $\mu_F, \mu_R$ set to $H_T/2$. The calculation includes resonant $t \bar t$ diagrams, single top diagrams, and diagrams without top quarks at all. For the purposes of this example, given the importance of off-shell effects, we reconstruct the observables $m_{b \bar b 4 \ell}$ and $D^{(1)}$ from the external particles without any acceptance restrictions that may stem from top-quark reconstruction algorithms. The toponium signal is simulated in {\tt MadGraph5\_aMC@NLO} with \eqref{httlagrangian} at LO (one-loop) in a separate sample with respect to the one containing the background, and added to it after generation. Note that there is no SM-$\phi$ interference in this case, as it is assumed that $t \bar t$ produced exactly at threshold are characterised by a Coulombic wave function which includes an all order effect resummed in the contribution from the scalar. Finally, note that the $\phi$ contribution is meant to represent the resummation of the terms at $\beta=0$ (and in fact slightly below threshold), so there is no double counting with the open quark singlet production in the LO QCD computation at $\beta>0$, which we include separately.

Without attempting a fit, which would be inappropriate given the current scarcity of experimental data and the large modelling uncertainty, we simply note that the tension between the recent ATLAS measurement of $D^{(1)}$ \cite{ATLAS:2023fsd} and the reference SM prediction seems to follow the pattern seen in Figure \ref{fig:line_threshold_D}. The overall size of the toponium contribution is given by the $t \, \bar t \, \phi$ coupling $c_y$, while the proportion between on-shell and off-shell effects is driven by the toponium width $\Gamma$.

\subsection{SUSY in the top-quark corridor}\label{sec:susy}

\begin{figure}[t]
    \centering
    \includegraphics[width=.8\textwidth]{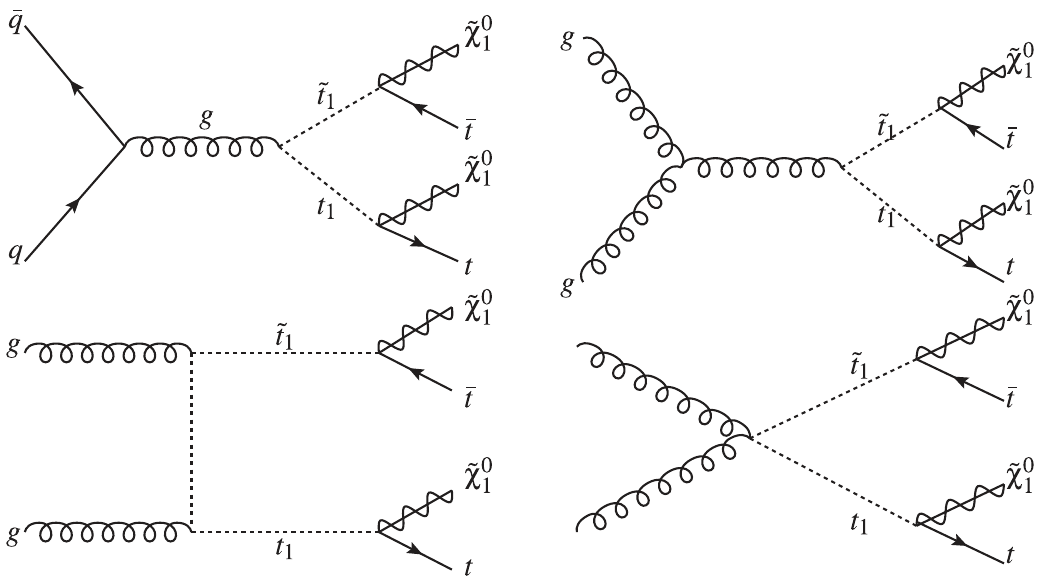}
    \caption{Representative Feynman diagrams describing production of a top squark pair, with the top squarks subsequently decaying into a (possibly off-shell) top quarks and neutralinos.}
    \label{fig:susydiags}
\end{figure}

Another interesting scenario is provided by pair production of supersymmetric top squarks decaying into top quarks and (stable) neutralinos, see Figure \ref{fig:susydiags}.

It is important to note that in this scenario the top squark decay chains,
\begin{align}
    \tilde{t}_1 \, \tilde{t}_1 \to t \, \bar t  \, \tilde{\chi}^0_1 \, \tilde{\chi}^0_1 \to b \, \bar b \, W^+ \, W^- \, \tilde{\chi}^0_1 \, \tilde{\chi}^0_1 , 
\end{align}
always contain a top quark pair, so that the interpretation of a measurement on final-state leptons in terms of SM top spin polarization/correlation/entanglement is still valid for top squark pair production events.

The region of parameter space we will consider is::
\begin{equation}
    |m_{\tilde{t}_1} - \, m_{\tilde{\chi}^0_1} - m_t| \approx \Gamma_t, \label{topcorridor}
\end{equation}
a region usually named the "{\it top mass corridor}". In this scenario the decay channel $\tilde{t}_1 \to t \, \tilde{\chi}^0_1$ (with the top quark possibly slightly off-shell) is expected to be dominant under most SUSY scenarios and results in a  top-quark pair with basically no additional missing momentum, in a configuration that is very similar of those produced by SM $t \bar t$ processes. This scenario is therefore very challenging and stop mass limits are typically weaker than outside the corridor. This challenge has been picked up by several groups in the last years suggesting new observables and techniques, see e.g.~\cite{Bagnaschi:2023cxg} for a recent proposal.

The scalar nature of top squarks yields significantly different spin correlations with respect to the SM background. Since the fermion lines of the SM top and anti-top quarks are disconnected and their spins uncorrelated,
\begin{equation}
    \mathcal C^{[\text{\tiny SUSY}]} = \begin{pmatrix} 0 & 0 & 0 \\ 0 & 0 & 0 \\ 0 & 0 & 0 \end{pmatrix}.
\end{equation}
As a result, 
\begin{equation}
    \mathcal C = \frac{A^{[\text{\tiny SM}]}}{A^{[\text{\tiny TOT}]}} \mathcal C^{[\text{\tiny SM}]} + \frac{A^{[\text{\tiny SUSY}]}}{A^{[\text{\tiny TOT}]}} \mathcal C^{[\text{\tiny SUSY}]} = \frac{A^{[\text{\tiny SM}]}}{A^{[\text{\tiny TOT}]}} \mathcal C^{[\text{\tiny SM}]},
\end{equation}
and clearly, since in the presence of a signal one has $A^{[\text{\tiny TOT}]} > A^{[\text{\tiny SM}]}$, the SUSY signature is a dilution of the SM spin correlations. 

While we focus here on spin correlations, it is interesting to note that the chiral nature of the $(\tilde{t}_1 \, t \, \tilde{\chi}^0_1)$ vertex violates parity and produces top quarks with large, order one, individual polarisation \cite{1811.08573}:
\begin{equation}
    B_{i} = \frac{A^{[\text{\tiny SM}]}}{A^{[\text{\tiny TOT}]}} B_{i}^{[\text{\tiny SM}]} + \frac{A^{[\text{\tiny SUSY}]}}{A^{[\text{\tiny TOT}]}} B_{i}^{[\text{\tiny SUSY}]} \simeq \frac{A^{[\text{\tiny SUSY}]}}{A^{[\text{\tiny TOT}]}} B_{i}^{[\text{\tiny SUSY}]},
\end{equation}
so that, since $A^{[\text{\tiny SUSY}]} > 0$, spin polarizations would be seen in the presence of a SUSY signal. (In the SM top polarizations arise only due to parity-violating EW interactions and are expected to be at the undetectable $10^{-3}$ level \cite{Bernreuther:2015yna}.)  

\section{Searches for new physics using quantum observables} \label{sec:analysis}

 We now investigate the sensitivity of quantum observables to resonant new physics at the LHC, for the two NP scenarios we described in Sections \ref{sec:scalar} and \ref{sec:susy}. We compare two classes of $t \bar t$ observables, ``classical'' observables, such as the differential distribution $d\sigma/dm_{t \bar t}$, and ``quantum'' observables, such as the amount of $t \bar t$ spin entanglement quantified by the $D$'s of \eqref{D1}-\eqref{D4}. 
 
We simulate a measurement with $140 \, \text{fb}^{-1}$ of luminosity of several entanglement and spin correlations-related observables, obtained differentially in the invariant mass of the $t \bar t$ pair decaying in the dilepton final state,
\begin{equation}
    p \, p \to t \, \bar t \to b \, \bar b \, \ell^+ \, \ell^- \, \nu_\ell \, \bar \nu_\ell. \label{process}
\end{equation}
We only consider the $e^+ \mu^-$ or $e^- \mu^+$ final states to improve background rejection, and assume a $30 \%$ overall reconstruction efficiency, as in \cite{ATLAS:2023fsd} and other similar dilepton analyses. The observables we consider are:
\begin{enumerate}[itemsep=0pt]
    \item The total number of events $N$.
    \item The singlet and triplet entanglement markers $D^{(1)}$, $D^{(k)}$, $D^{(r)}$, and $D^{(n)}$, defined in \eqref{D1}-\eqref{D4}.
    \item The average angular separations of leptons, $\Delta \eta = |\eta_{\ell^+} - \eta_{\ell^-}|$ and $\Delta \phi = |\phi_{\ell^+} - \phi_{\ell^-}|$.
    \item The average aperture between leptons in the laboratory frame $\cos \varphi = \hat{p_{\ell^+}} \cdot \hat{p_{\ell^-}}$.
\end{enumerate}
The feasibility of such measurements has already been demonstrated. The quantum observable $D^{(1)}$ has been measured inclusively using $39 \, \text{fb}^{-1}$ of data with an absolute uncertainty of $\pm 0.011$  \cite{CMS:2019nrx}, and differentially in the $t \bar t$ invariant mass using $140 \, \text{fb}^{-1}$ of luminosity, with a final absolute uncertainty of $\pm 0.021$ in the bin that yielded the first observation of entanglement \cite{ATLAS:2023fsd}. Most of the uncertainty in \cite{ATLAS:2023fsd} stems from signal modelling (and therefore is expected to improve in the future), with the second-largest source being from backgrounds. Experimental uncertainties associated to ISR or $b$-tagging, that may in principle be relevant for a $t \bar t$ dilepton analysis, have been found to be largely subleading.

It is important to note that the percent-level accuracy reached in \cite{ATLAS:2023fsd} is only possible due to the existence of a dedicated experimental handle, the opening angle of leptons, that is directly sensitive to $D^{(1)}$ without the need to reconstruct $C_{kk}$, $C_{rr}$, and $C_{nn}$ individually. In \cite{Aguilar-Saavedra:2022uye}, a similar quantity has been proposed for $D^{(k)}, D^{(k)}, D^{(n)}$, which may eventually yield a similar experimental sensitivity. The observables $\Delta \eta$, $\Delta \phi$, and $\cos \varphi$ are not purely sensitive spin correlations, but rather to a combination of spin and kinematics. Nevertheless, the excellent experimental reconstruction capability of the unboosted lepton momenta prompts us to include them in our analysis. 

We assume statistical and systematic uncertainties on our observables as listed in Table \ref{tab:unc}. 
A systematic uncertainty from finite detector resolution is assigned on $m_{t \bar t}$ and on all observables based on existing experimental measurements, as well as a Poisson statistical uncertainty of the form $\Delta x = x_0 / \sqrt{N}$, where $x_0$ is a typical value of the observable $x$ and $N$ is the expected number of events in each bin at the detector level, after the branching fraction and acceptance cuts have been considered.

\begin{table}[t]
\centering
\begin{tabular}{|c c c|} 
 \hline
 Observable & Systematic unc. & Statistical unc.\\
 \hline\hline
  $m_{t \bar t}$ & $30 \ \text{GeV}$ & \\ 
  \hline\hline
 $\nicefrac{dN}{dm_{t \bar t}}$ & $0.03 \, \cdot \, N$ & $\sqrt{N}$ \\ 
  \hline
 $\nicefrac{d(\cos \varphi)}{dm_{t \bar t}}$ & 0.010 & $0.5 / \sqrt{N}$\\
 \hline
  $\nicefrac{d(\Delta \eta)}{/dm_{t \bar t}}$ & 0.010 & $3 / \sqrt{N}$\\
 \hline
 $\nicefrac{d(\Delta \phi)}{dm_{t \bar t}}$ & 0.010 & $2.5 / \sqrt{N}$\\
 \hline
 $\nicefrac{dD^{(1)}}{dm_{t \bar t}}$ & 0.015 & $0.75 / \sqrt{N}$ \\
  \hline
 $\nicefrac{dD^{(k,r,n)}}{dm_{t \bar t}}$ & 0.025 & $0.75 / \sqrt{N}$\\
 \hline
\end{tabular}
\caption{Estimated absolute experimental uncertainty at the LHC on the observables considered in our study, obtained by comparison with similar measurements \cite{CMS:2018rkg, CMS:2019nrx, 2005.05138, ATLAS:2023fsd, 2303.15340}, and with simulations.}
\label{tab:unc}
\end{table}

The analyses presented in this work are based on simulations with {\tt MadGraph5\_aMC@NLO} \cite{1405.0301}. Proton-proton collisions are simulated at  $\sqrt{s} = 13 \, \text{TeV}$ using the {\tt NNPDF4.0} parton distribution functions \cite{NNPDF:2021njg}. 
The renormalisation and factorisation scales are set to the dynamical value $\mu_R = \mu_F = H_T/2$. The SM background contains all tree-level diagrams for the process \eqref{process}, that is, diagrams at order $\alpha_{s}^2 \,  \alpha^4$, corresponding to QCD top production, as well as diagrams with electroweak vertices, at order $\alpha_{s} \, \alpha^5$, and $\alpha^6$.
As described in more detail in each of the following sections, the new physics signal is generated from suitable {\tt UFO} models at LO, and added on the SM background, including the corresponding interference if relevant. We do not anticipate the restriction to LO accuracy to significantly alter our  results. It is known that higher order QCD corrections to spin observables are not sizable in the SM \cite{1508.05271, Czakon:2020qbd} and in most NP scenarios \cite{2210.09330}, and since spin observables are defined as ratios of (correlated) cross-sections, they carry a significantly smaller theoretical uncertainty than the individual ingredients used to construct them. Furthermore, our search is for resonant particles, and therefore amounts to bump--hunting, which does not require an exceedingly accurate SM background prediction.

\subsection{Scalar/pseudoscalar resonances}\label{sec:scalar_analysis}

\begin{figure}[p]
    \centering
    \includegraphics[width=0.49\textwidth]{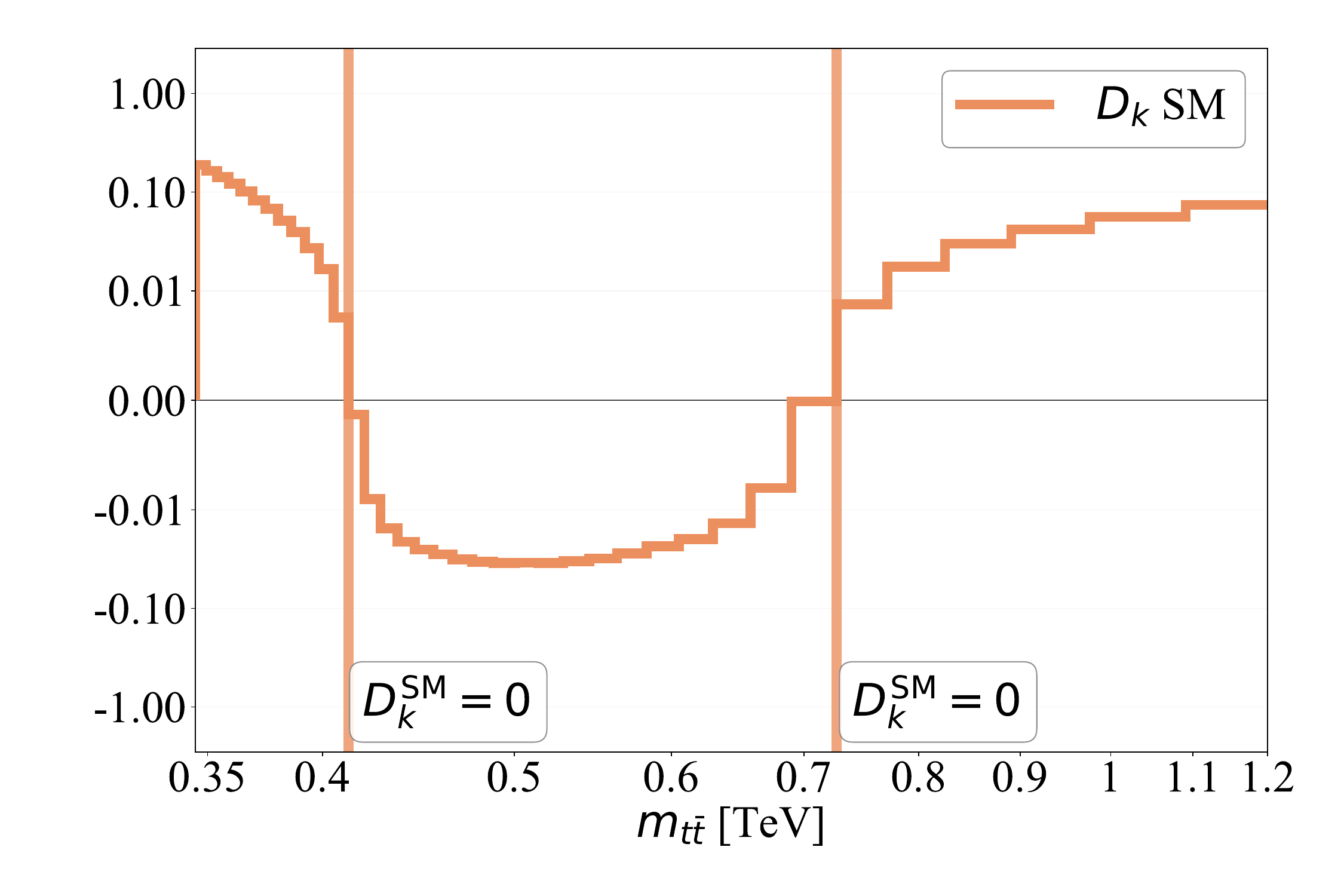}
    \includegraphics[width=0.49\textwidth]{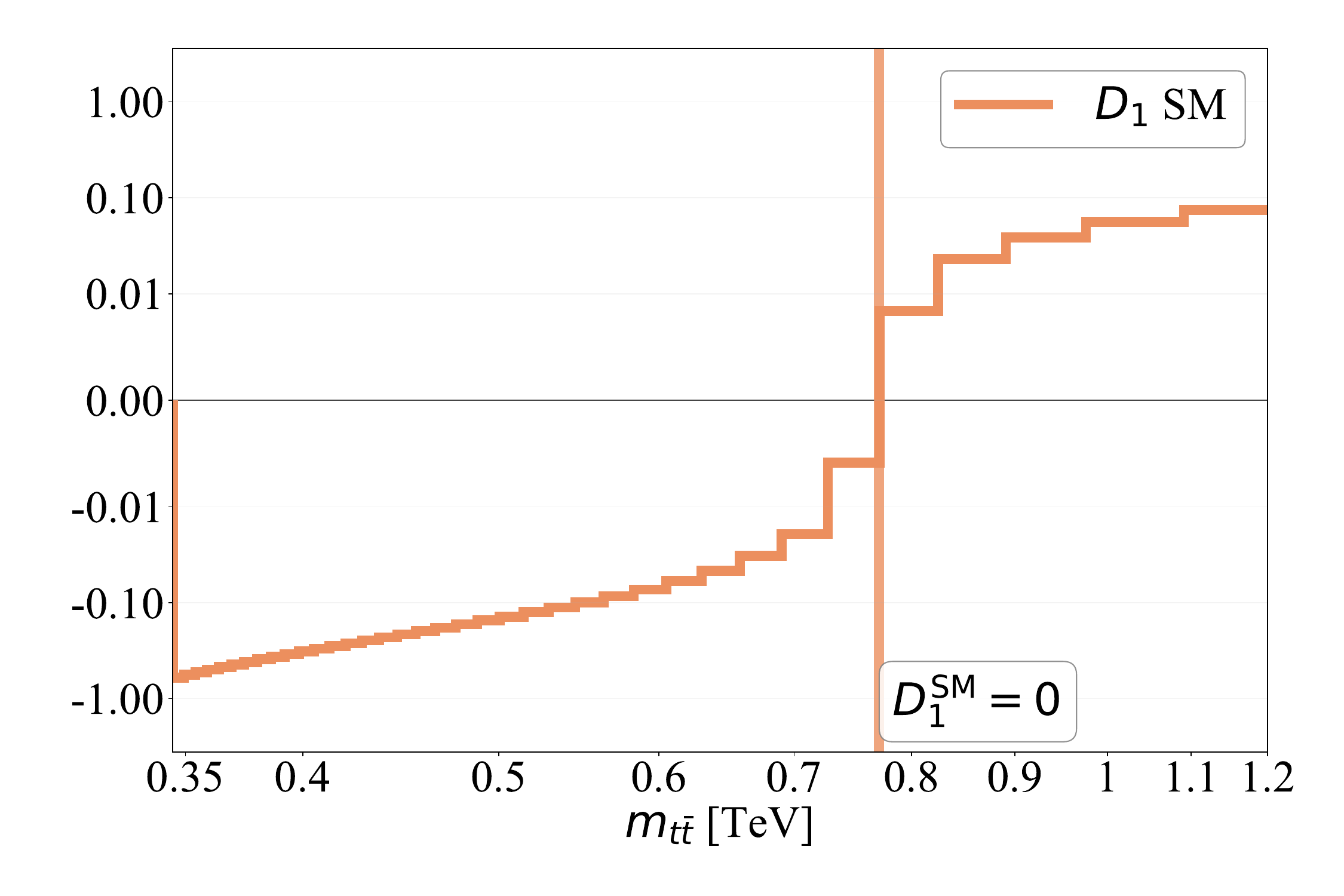}
    \includegraphics[width=0.49\textwidth]{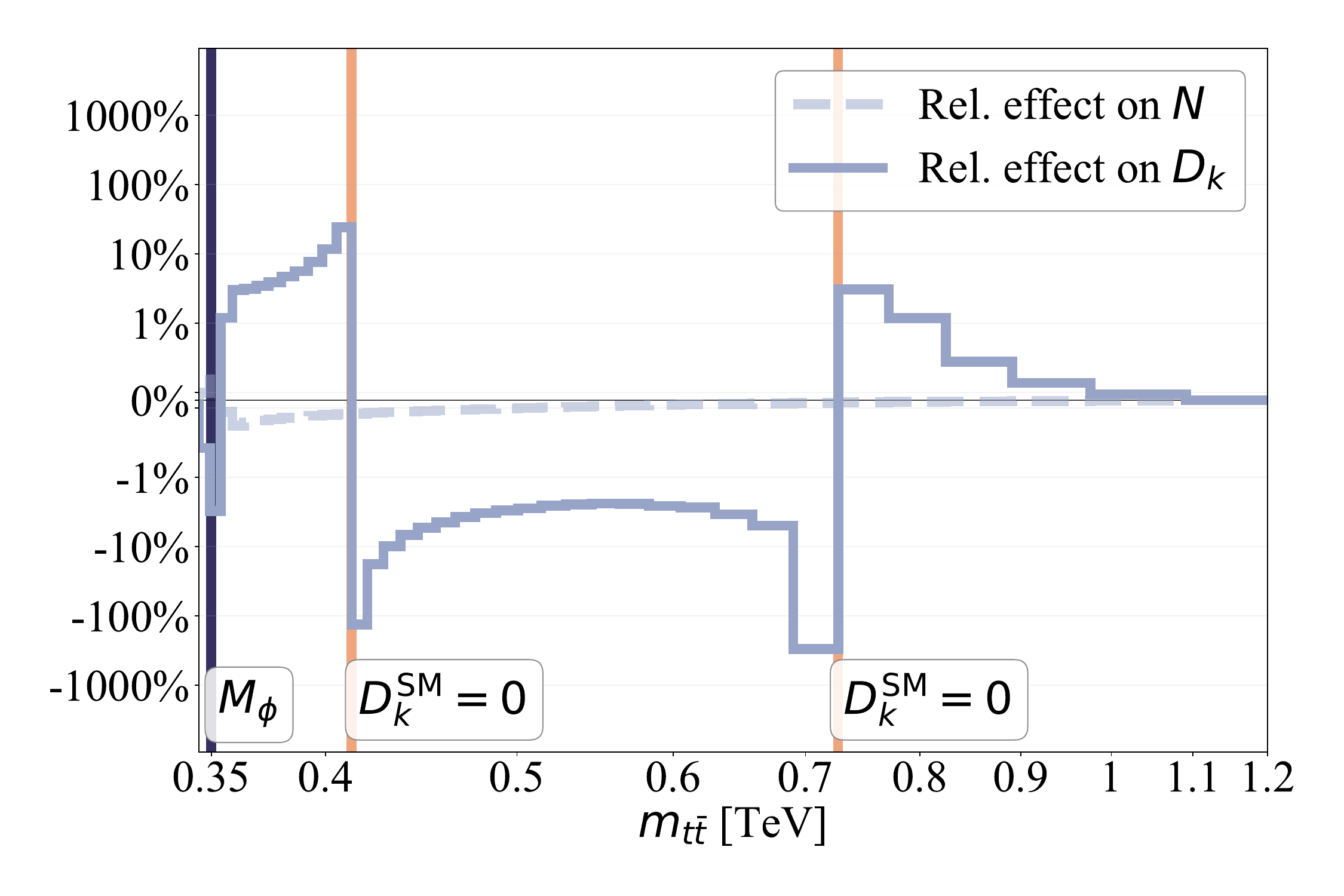}
    \includegraphics[width=0.49\textwidth]{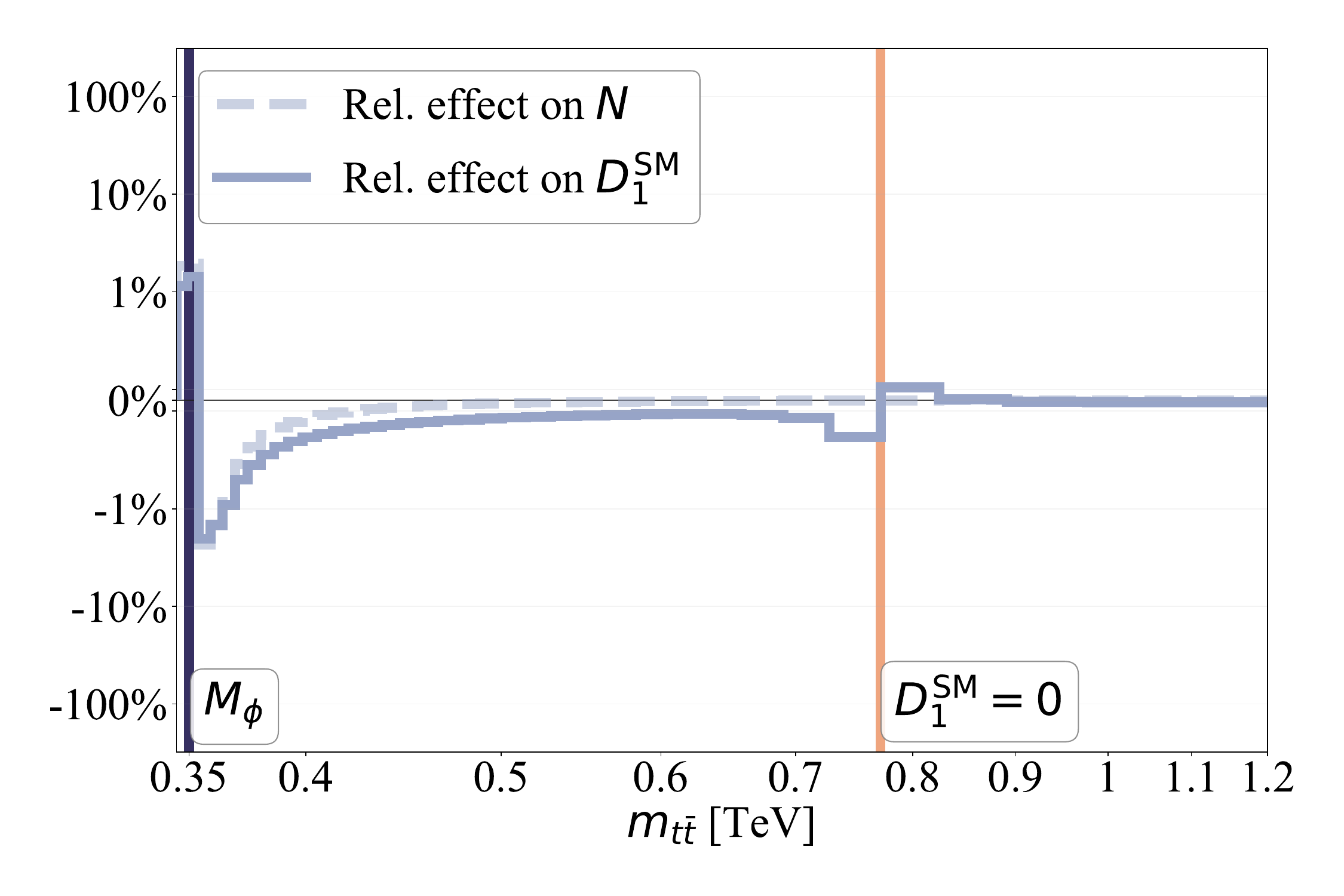}
    \includegraphics[width=0.49\textwidth]{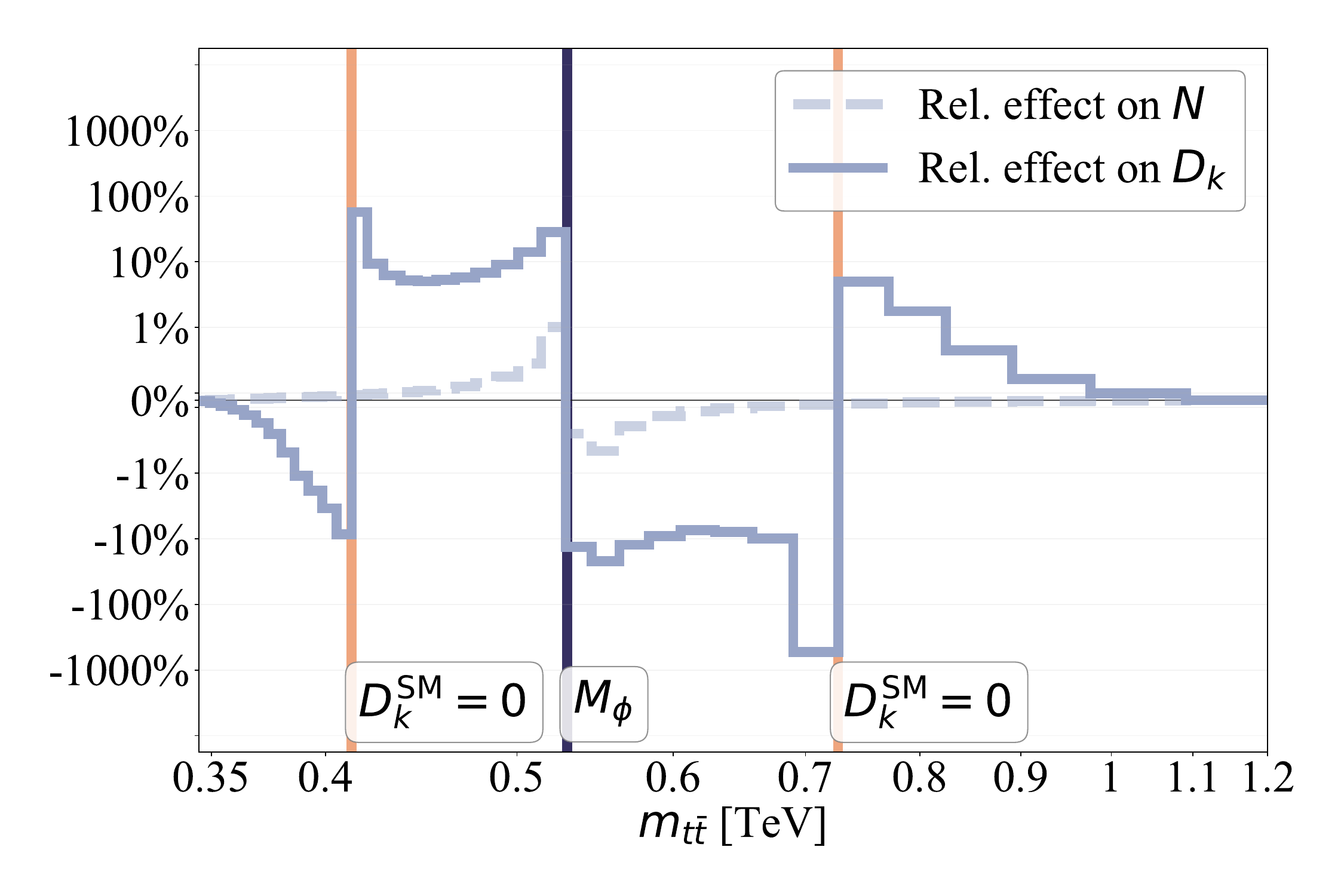}
    \includegraphics[width=0.49\textwidth]{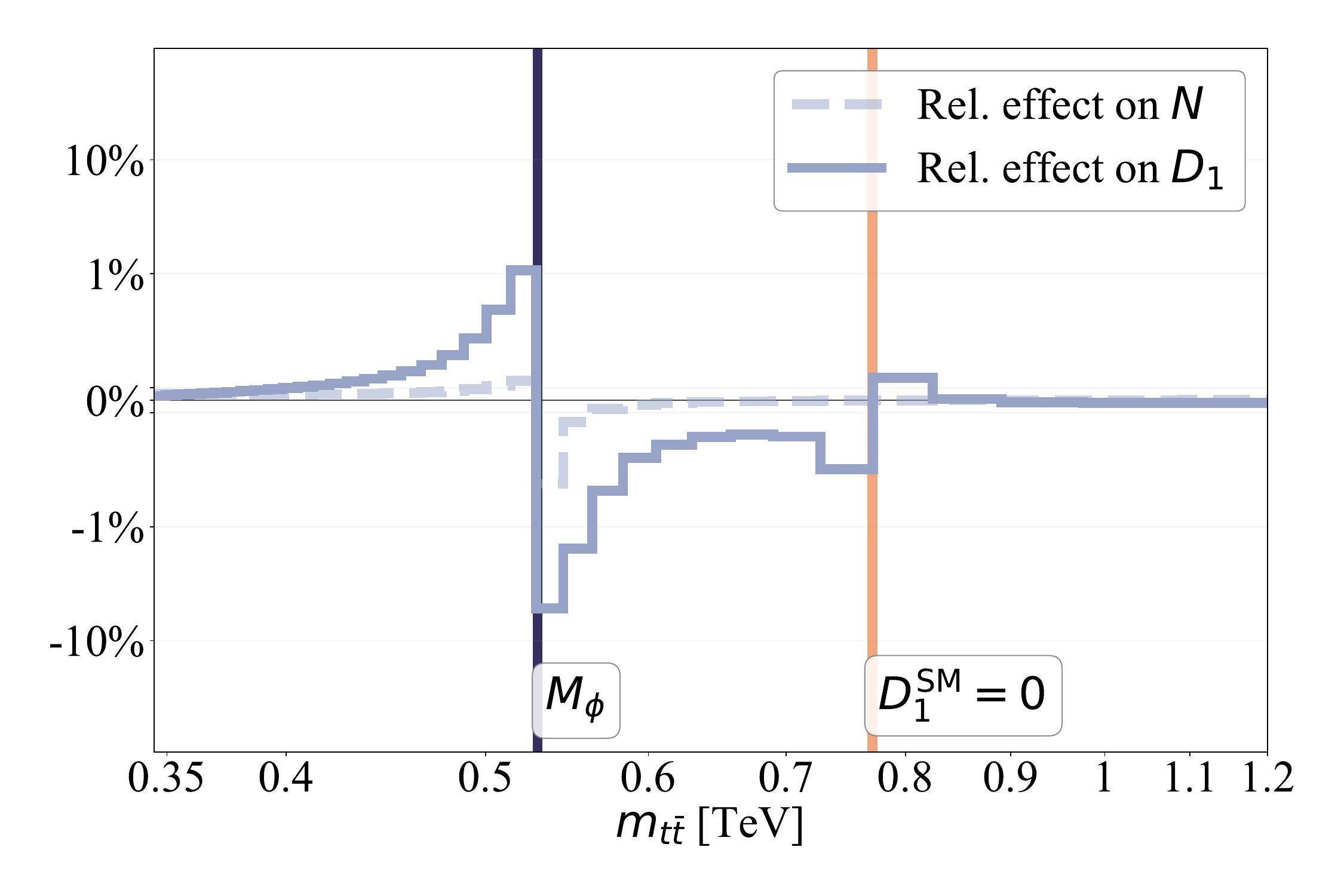}
    \includegraphics[width=0.49\textwidth]{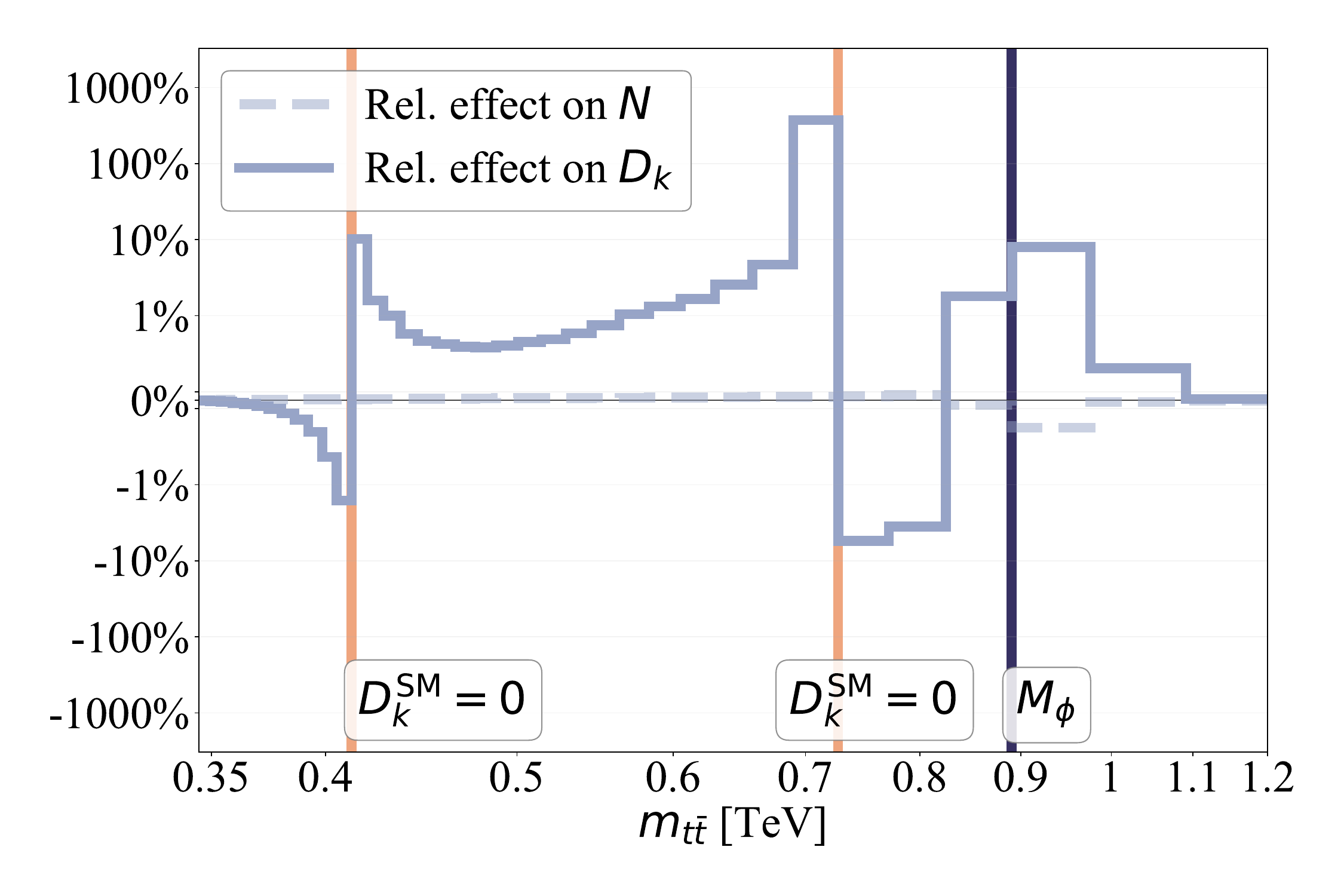}
    \includegraphics[width=0.49\textwidth]{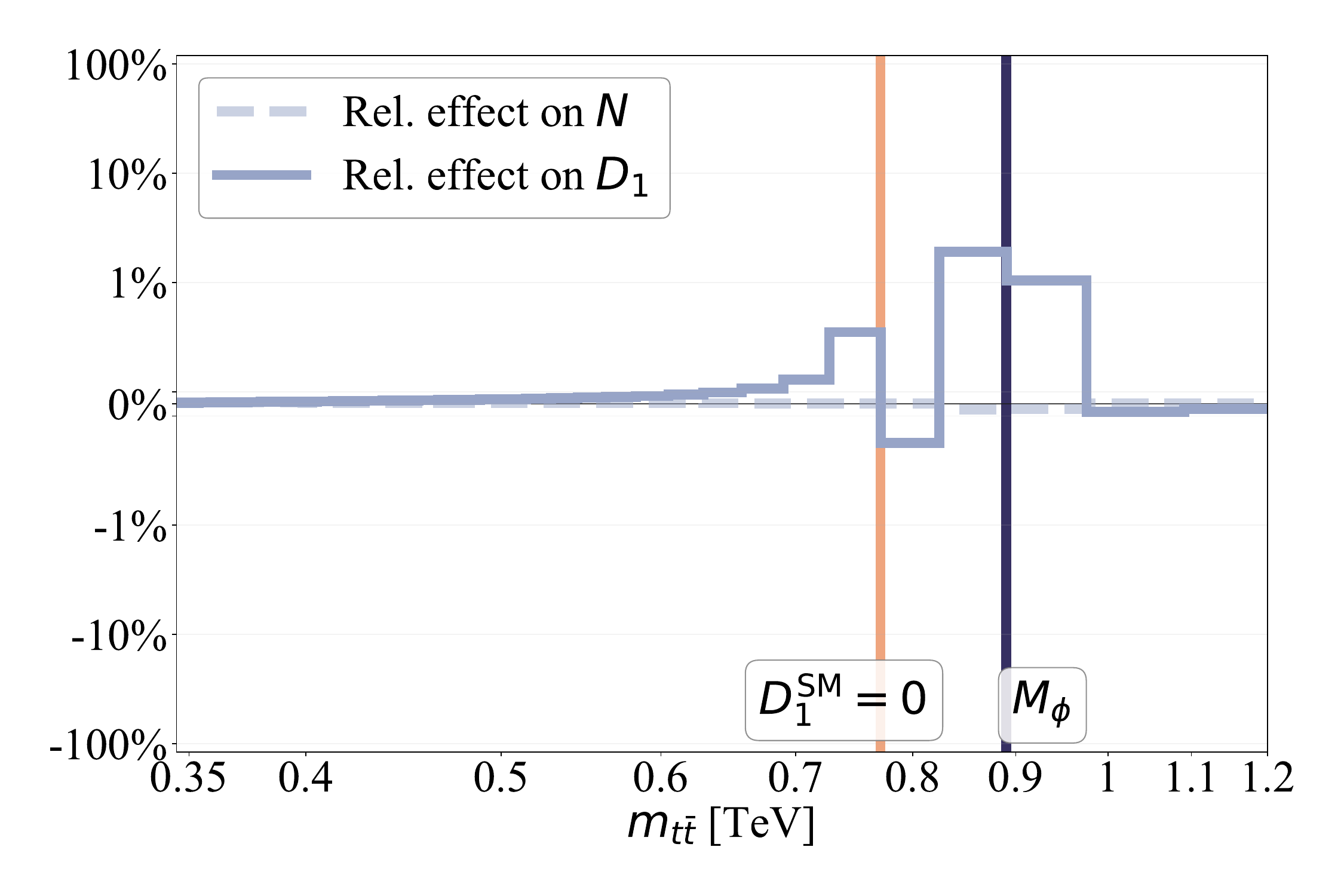}
    \caption{Effect of $\phi$ on the distributions $dN / dm_{t \bar t}$ (solid) and $dD / dm_{t \bar t}$ (dashed). Top: SM distributions. Second, third, and fourth rows: scalar (left) and pseudoscalar (right) relative contribution to the SM, for $c_y = 0.5$. From top to bottom: $M = 0.35$, $0.53$, $0.89$ TeV. For the scalar case the entanglement marker is $D^{(k)}$, while for the pseudoscalar it is $D^{(1)}$. Vertical lines show the values of $m_{t \bar t}$ for which the SM entanglement markers are zero and the mass of the (pseudo)scalar. 
    }
    \label{fig:mtt_distribution}
\end{figure}

We first consider the case of the production of a heavy scalar, with CP-even and CP-odd couplings to the top quark, as introduced in Section \ref{sec:scalar}. This scenario is a particularly favourable playground for testing the sensitivity of quantum observables, as already explored, for instance, by the CMS collaboration \cite{CMS:2019pzc}. We show several examples of the effect of the heavy scalar on the distributions $dN / dm_{t \bar t}$ and $d D / dm_{t \bar t}$ in Figure \ref{fig:mtt_distribution}. As evident from the plots, a value of the (pseudo)scalar mass much larger than $2 m_{t}$ quickly suppresses the NP contribution to $dN / dm_{t \bar t}$, resulting in a likely invisible sub-\% effect. On the other hand, the effect on $D$ is significantly larger, being consistently of order $\gtrsim$ 10\% across the $m_{t \bar t}$ range we analysed, with spikes of 100\% or more where the SM value for $D$ nears zero, and close to the $\phi$ resonance.

\begin{figure}[t]
    \centering
    \includegraphics[width=0.6\textwidth]{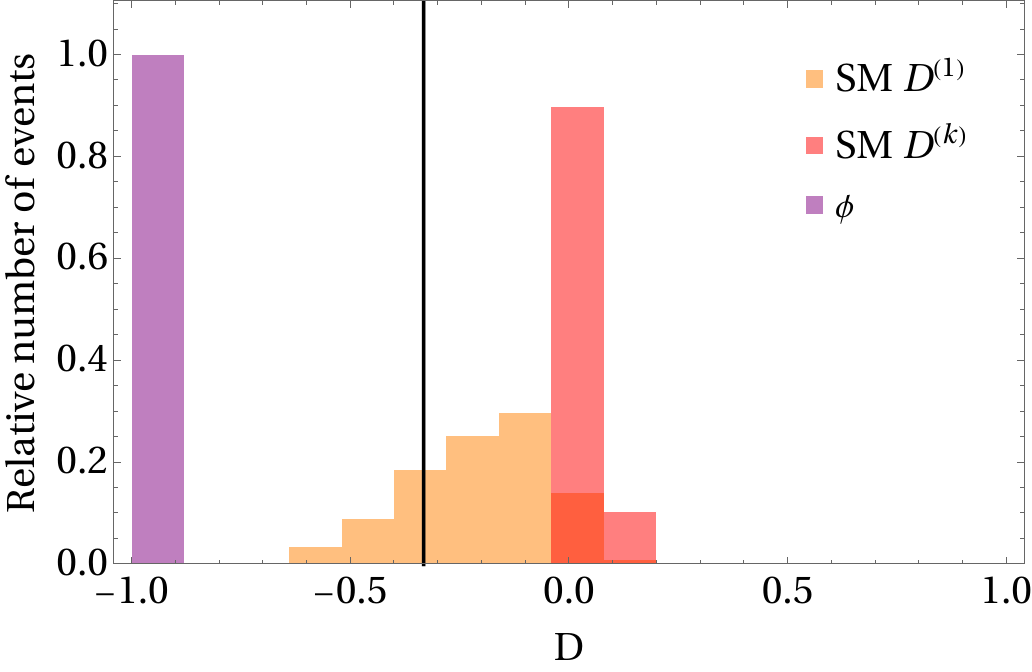}
    \caption{Relative number of events, $N_{\text{bin}} / N_{\text{tot}}$, in each range of values $[D, D+dD]$ for the SM (orange and red) and for the interference and signal for the heavy scalar $\phi$ (purple). A vertical line at $D=-\nicefrac 1 3$ divides the classical from the entangled region. }
    \label{fig:ddistr}
\end{figure}

The significant difference in the values of $D$ between the SM and the heavy scalar $\phi$ becomes even more impressive when events are collected in regions of phase space with a similar value of $D$. We show the result of such grouping in Figure \ref{fig:ddistr} for the SM, for the $\phi$-SM interference, and $\phi$-squared channels.

For our simulated search we have prepared a suitable {\tt UFO} model and, to ensure the possibility of treating $\phi$ as a relatively narrow resonance, imposed the condition:
\begin{equation}
   \Gamma \,\le \, \frac{M_\phi}{2}, \label{eq:eq_widthlimit}
\end{equation}
 that identifies an upper bound for the $c_y$ parameter for each value of the resonance mass. Our simulation includes purely-signal contributions, and all interferences between the signal amplitude and the SM QCD background. The signal amplitude is taken to be the diagram in Figure \ref{fig:htt}, which is highly dominant with respect to all other (non--$s$-channel) contributions at the same order. The width of $\phi$ is taken from its LO expression \eqref{htt_width}. While $\phi$ may decay in SM states other than $t \bar t$, we will assume that when $M_\phi \gg 2m_t$ we have $\Gamma = \Gamma_{\phi \rightarrow t\bar t}$, while near $M_\phi  = 2 m_t$ (where other $\phi$ decay channels become dominant) we take $\Gamma = 1.5 \, \text{GeV}$.

\begin{figure}[t]
    \centering
    \includegraphics[width=0.7\textwidth]{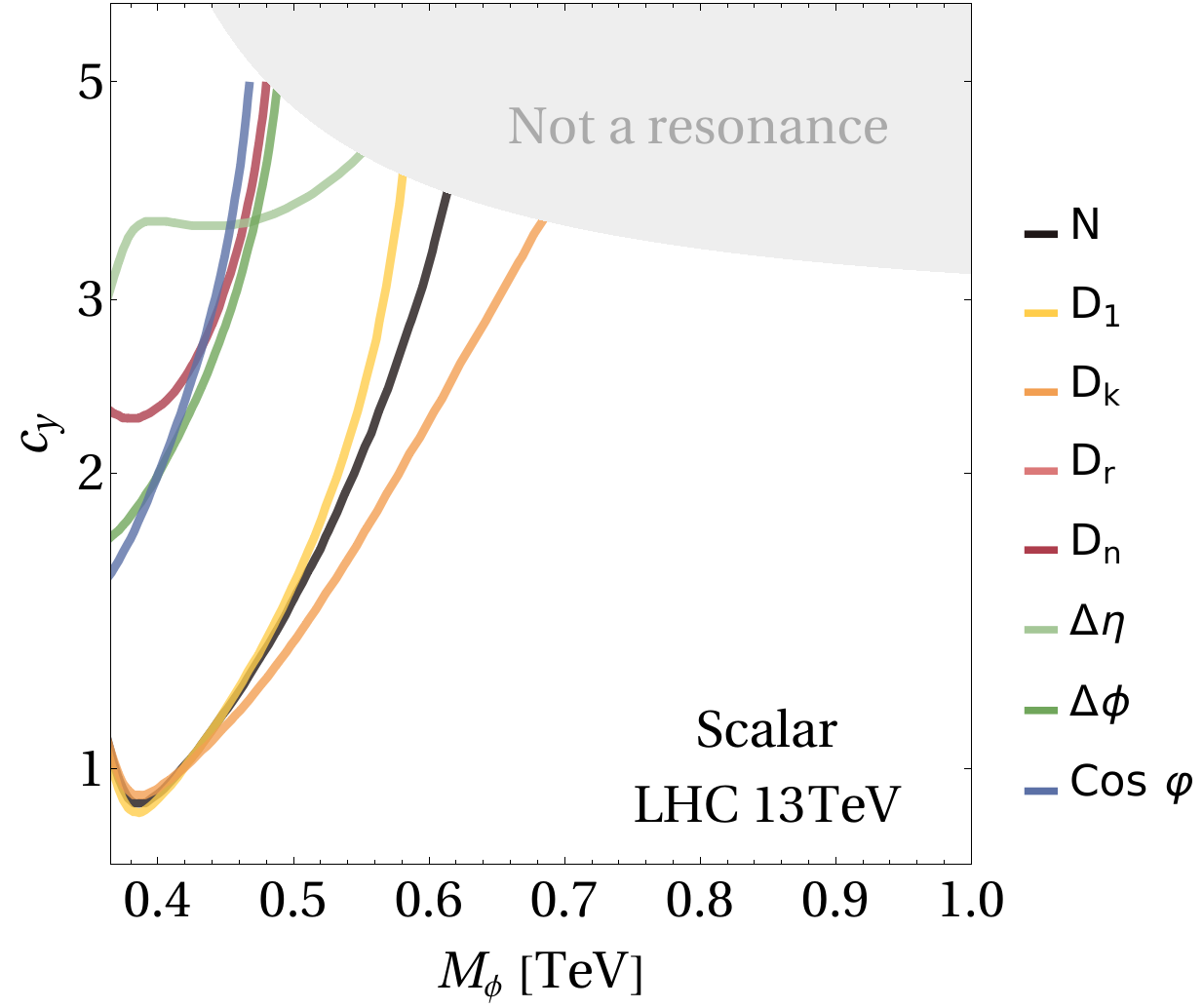}
    \caption{Results of our simulated search in $(M_\phi,c_y)$ parameter space for $\alpha = 0$ using one observable at a time. Regions yielding a visible signal (at $95 \ \%$ local significance) are above and to the left of the plotted bounds, while those that can not be excluded by this search are below and to the right. The region where \eqref{eq:eq_widthlimit} is not satisfied is shaded in grey.}
    \label{fig:scalar}
\end{figure}

\begin{figure}[t]
    \centering
    \includegraphics[width=0.7\textwidth]{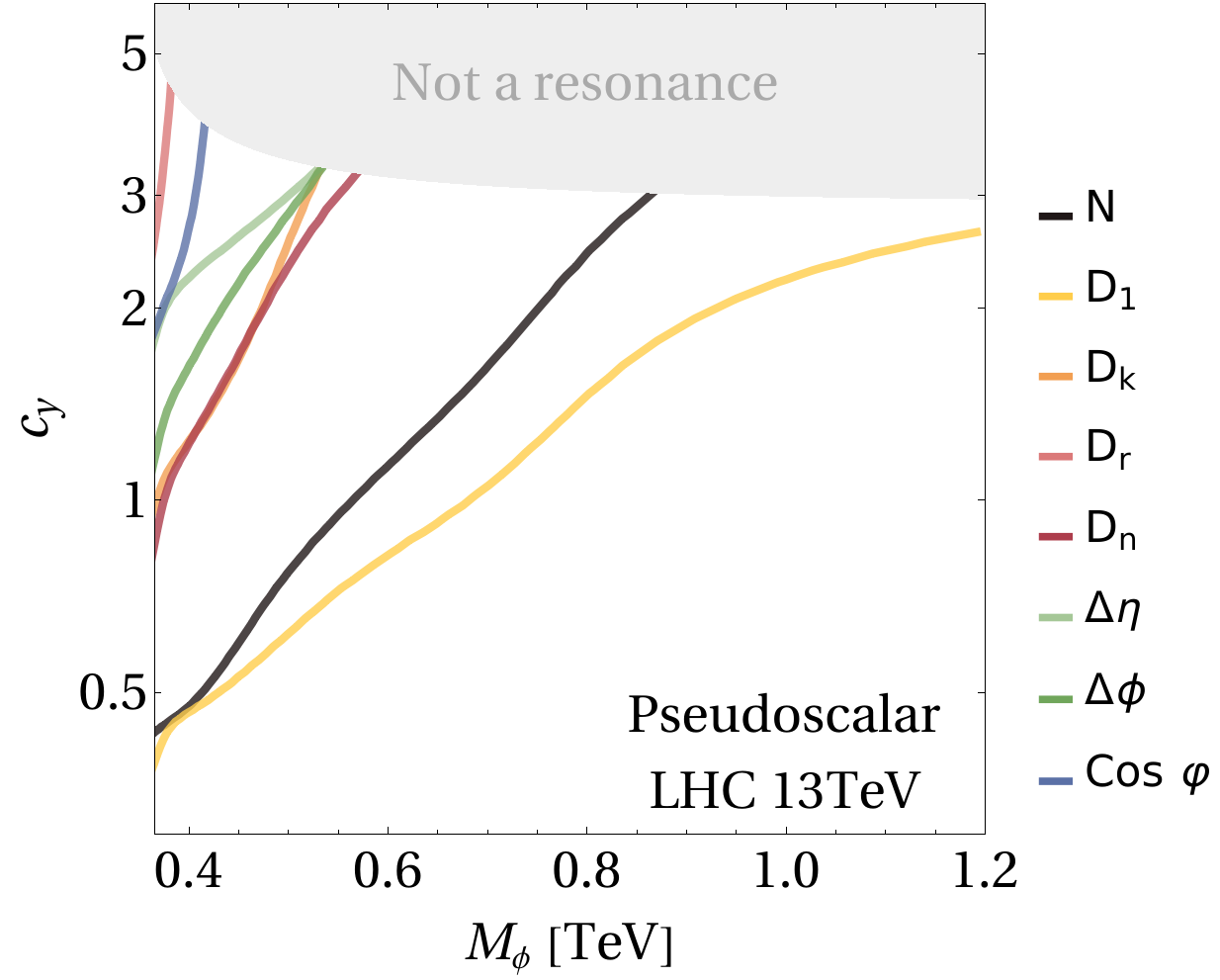}
    \caption{Results of our simulated search in $(M_\phi,c_y)$ parameter space for $\alpha = \pi/2$, one observable at a time, similar to Figure \ref{fig:scalar}. (Note the different axes range.)}
    \label{fig:pseudoscalar}
\end{figure}

The results from our full simulated analysis are shown in Figure \ref{fig:scalar} and \ref{fig:pseudoscalar}, where the regions of parameter space excluded at $95 \, \%$ local significance are shown for the scalar and pseudoscalar case.

The shapes of the exclusion regions we obtain are dictated by the interplay between the mass $M_\phi$, width $\Gamma_\phi$, and coupling strength $c_y$ of $\phi$, but a few interesting general facts may be noted. As it can be seen from the Figures, there are multiple regions of parameter space that produce a visible effect in $t\bar t$ spin entanglement, but are invisible with more conventional searches. In particular, the most sensitive entanglement observable for the scalar resonance is $D^{(k)}$, while the most sensitive entanglement observable for the pseudoscalar is $D^{(1)}$, as expected from the respective quantum states \eqref{H}.

The case of a resonance near threshold is interesting especially for $\alpha = \pi/2$, as our model may be taken as a simplified description of $t \bar t$ bound states, as discussed in Section \ref{sec:toponium}. We note that the greatest sensitivity at threshold is given by $D^{(1)}$ and by the total rate $N$. It is therefore conceivable that, if the large down-fluctuation observed by ATLAS in $D^{(1)}$ near threshold is indeed a pseudoscalar resonance, a similarly sized up-fluctuation would also be present in the cross-section, and it is unlikely that such a signal will be visible anywhere else.

While the exact location of the exclusion boundaries depend on our uncertainty estimates, we find that, unless the experimental resolution on $D$ will turn out to be significantly worse than anticipated, searches for $\phi$ based on spin observables consistently prove to be better than searches based on kinematical distributions. We have checked that this important conclusion is largely independent of our uncertainty estimates.

\subsection{Heavy neutral boson} \label{sec:zprime}

Another interesting BSM scenario is given by the resonant production of a heavy vector boson $Z'$, similar to the SM $Z$ but with multi-TeV mass, so that the decay into an on-shell top pair is possible. We have extracted the spin correlation coefficients of top/anti-top pairs produced by a vector boson with arbitrary couplings in Section \ref{sec:sm_ew}. 

As an example, we consider a heavy $Z'$ boson that couples to all quarks with an interaction with the same couplings and chiral structure as the SM $Z$ boson. In our analysis we take the $Z'$ mass to be $m_{Z'} > 2 \, \text{TeV}$ and its width to mass ratio to be $\Gamma_{Z'} / m_{Z'} = 3 \, \%$, similarly to the SM electroweak bosons. 

Unlike the scalar/pseudoscalar case we considered above, that was largely limited by systematics, the search for such scenario is still limited by the statistical uncertainty. For our simulated search we take an integrated luminosity of $140 \, \text{fb}^{-1}$, corresponding to the data already on tape from Run 1 and Run 2. Of course, the amount of data available for such studies is expected to grow significantly in the upcoming years.

\begin{figure}[t]
    \centering\includegraphics[width=0.7\textwidth]{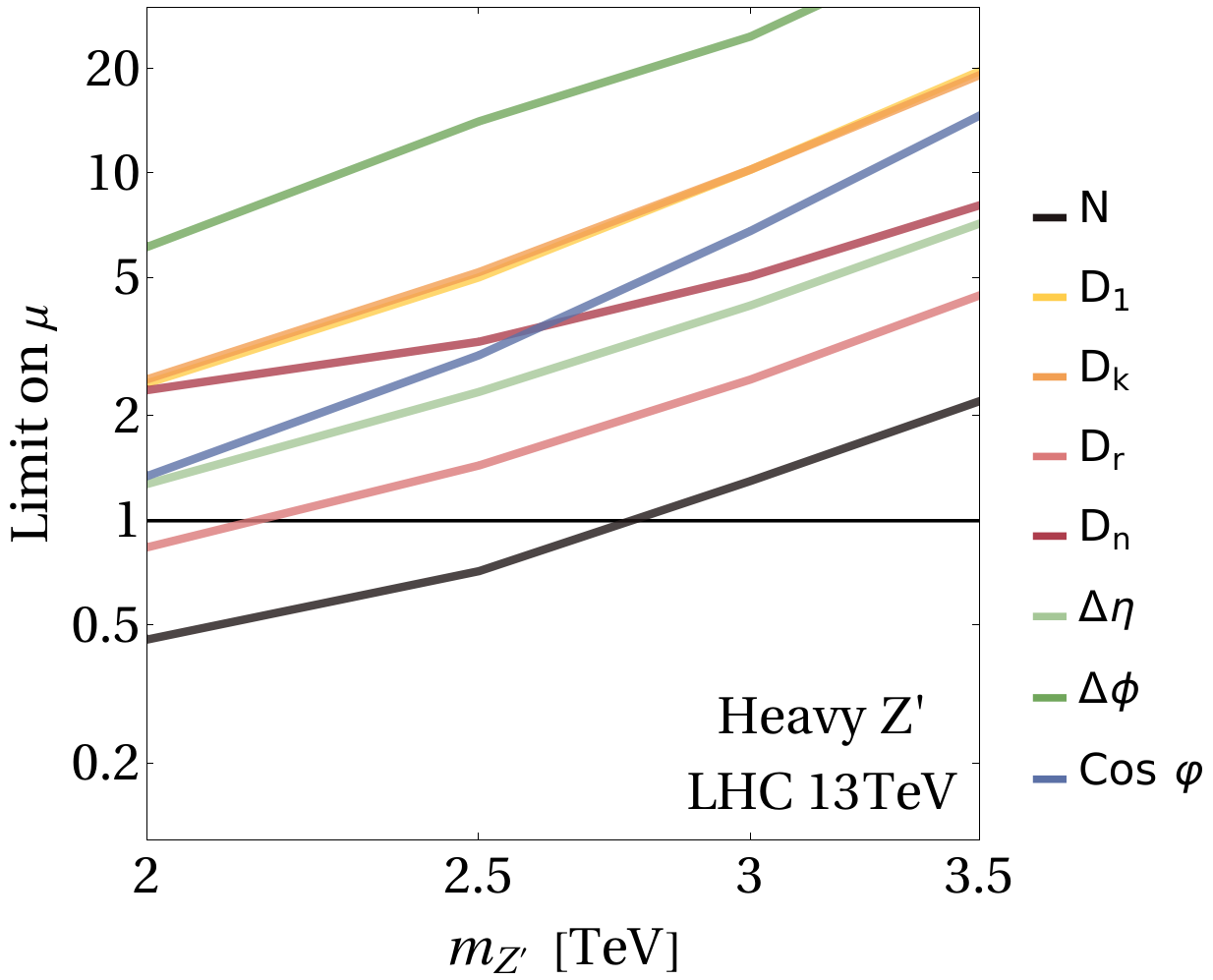}
    \caption{Results of our simulated search for a $Z'$, using one observable at a time. Limits are set on the signal strength $\mu$ at $95 \ \%$ local significance, as described in the text.}
    \label{fig:zprime}
\end{figure}

Figure \ref{fig:zprime} shows the exclusion ranges we obtain for all the observables we considered, one at a time. The $Z'$ signal is rescaled by a factor $\mu$, and we show the expected exclusion region in $\mu$ at $95 \%$ local significance. (For instance, an exclusion boundary of $\mu = 5$ at $m_{Z'} = 3 \, \text{TeV}$ means that to observe/exclude a $Z'$ of mass $3 \, \text{TeV}$ at $95 \%$ CL, its production cross section would have to be $5$ times larger than the actual prediction.)

Unlike the scalar/pseudoscalar particle, in this particular case top spin and entanglement observables are not expected to improve the bounds obtained by a simple bump-hunting in the total number of events. This is because the $t \bar t$ spin state reached by a vector resonance has a similar structure to the QCD background. Namely, for $\beta \to 1$, corresponding to a heavy intermediate resonance, we find:
\begin{equation}
    C_{kk} = 1, \qquad  C_{rr} = - C_{nn},
\end{equation}
and all other entries zero, for QCD as well as for all electroweak channels \eqref{smaqq1}-\eqref{smavqqn}.

\subsection{SUSY in the top-quark corridor}

A rather different situation arises in the case of production of top quarks stemming from the decay of their supersymmetric partners, especially when masses lie in the corridor $m_{\tilde{t}_1} - \, m_{\tilde{\chi}^0_1} \simeq m_t$. Similarly to the case we considered in Section \ref{sec:scalar_analysis}, this scenario is hard to see in kinematical distributions, and in fact general purpose SUSY searches based only on kinematics \cite{1706.04402,1707.03316,1708.03247,1710.11188,1711.00752,1912.08887,2008.05936,2012.03799,2102.01444,2103.01290,ATLAS:2023dbq} generally yield worse constraints in the top-quark corridor than in other regions.

The addition of spin and entanglement observables may change this situation radically. As noted above, the signature of SUSY top-squark production is a reduction of top spin correlations (and therefore entanglement) and an increase in individual top/anti-top polarization. A measurement from the ATLAS Collaboration \cite{ATLAS:2019zrq} already exploited the SUSY effects in spin correlations to explore the top-quark mass corridor with early Run 2 data. A recent CMS projection \cite{CMS:2022nmv} simulated a search in the top mass corridor for HL-LHC, using the full $t \bar t$ spin density matrix and additional quantities related to angular separations \cite{Han:2012fw}. It was found that the discovery potential for light top squarks in the top-quark corridor may be improved by an order of magnitude in the stop mass thanks to spin observables and the larger statistics.

Building upon \cite{ATLAS:2019zrq, CMS:2022nmv}, we simulate a search for the supersymmetric scenario we described in Section \ref{sec:susy}, for the most challenging case of complete mass degeneracy,
\begin{equation}
    m_{\tilde{t}_1} = m_t + m_{\tilde{\chi}^0_1}. \label{stop_mass}
\end{equation}
We write the interaction Lagrangian as:
\begin{equation}
    \mathcal L \, \supset \, i \, \bar t \, (g_L P_L + g_R P_R) \, \tilde{\chi}_1^0 \, \tilde{t}_1 + \textit{h.c.}
\end{equation}
The channel relevant for $t \bar t$ dilepton searches is:
\begin{equation}
    p \, p \to \tilde{t}_1 \, {\tilde{t}}^*_1 \to b \, \bar b \, \ell^+ \, \ell^- \, \nu_\ell \, \bar \nu_\ell \, \tilde{\chi}^0_1 \, \tilde{\chi}^0_1, \label{susy_process}
\end{equation}
which overlaps with \eqref{process}. If the $(\tilde{t}_1 \, t \, \chi_0^1)$ coupling is of order of the standard model $(t \, b \, W)$ coupling, the top-squark partial width into $\tilde{\chi}^0_1 \, b \, W$ becomes of order MeV or less, the narrow width approximation (NWA) is valid. In the NWA the top squarks are always produced on-shell, and our results do not depend on $g_L$ and $g_R$, but only on the stop mass. We discuss generation details in more depth in Appendix \ref{app:susy_gen}.

\begin{figure}[t]
    \centering
    \includegraphics[width=0.7\textwidth]{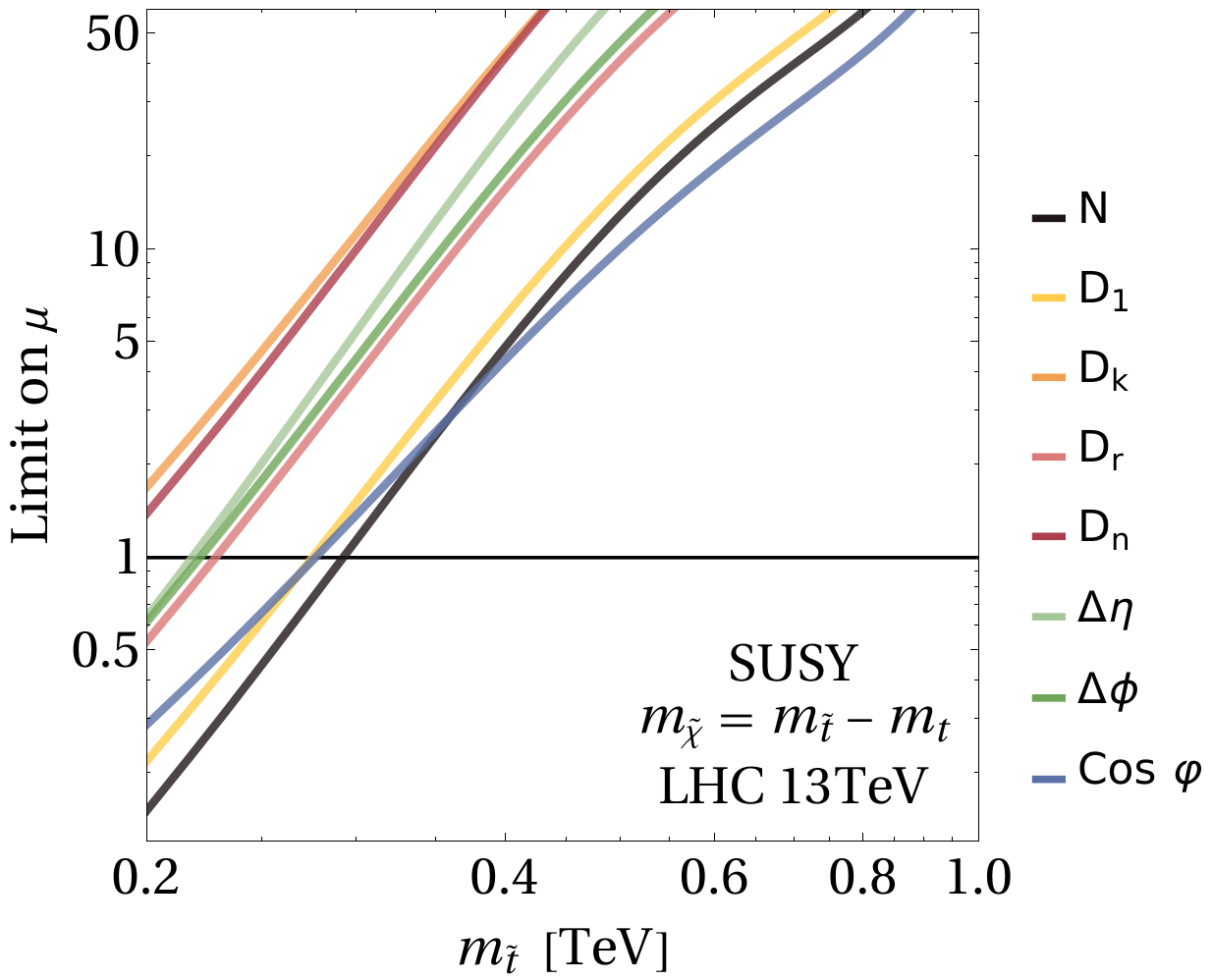}
    \caption{Results of our simulated search for top squarks and neutralinos in the top-quark corridor, using one observable at a time. Limits are set on the signal strength $\mu$ at $95 \ \%$ (local), similarly to figure \ref{fig:zprime}.}
    \label{fig:susy}
\end{figure}

As already noted above, the scalar nature of the top squarks  produces zero top spin correlations and order-one individual polarizations, a signature opposite to the QCD background, of no polarizations and significant spin correlations. The results of our simulated analysis are in Figure \ref{fig:susy}. The SUSY signal is rescaled by a strength parameter $\mu$ and limits on $\mu$ are shown for $95 \ \%$ local significance for a variety of top-squark masses $m_{\tilde t}$ and $m_{\tilde{\chi}^0_1} = m_{\tilde{t}_1} - m_t$.

Similarly to the scalar/pseudoscalar resonance studied in Section \ref{sec:scalar_analysis}, we find that some angular observables give an improved discovery power with respect to raw kinematical quantities, such as a bump in the number of events. Our analysis suggests the separation of leptons $\cos \varphi$ is the most sensitive observable for high stop masses, followed by $D^{(1)}$ and the total rate $N$. While the limits we obtain are not competitive with global combinations, our results show that the addition of spin and entanglement-inspired observables to searches for supersymmetric particles will improve bounds, perhaps significantly, in previously hard to exclude regions. 

While very promising, top spin measurements are not the only handle for detecting SUSY in the top-quark corridor, a dedicated analysis based on ISR and one based on the so-called $am_{T2}$ variable (an asymmetric transverse mass) also yielded encouraging results in Run 2 \cite{1709.04183,ATLAS:2017eoo,2004.14060}.

\section{Conclusions} \label{sec:conclusions}

In this work we have explored the sensitivity of quantum observables to new physics resonances in $t\bar t$ production.  We have computed the spin correlation matrix $\mathcal C$ analytically for all top-quark production channels relevant for the LHC. Besides re-deriving SM predictions, we have considered new physics scenarios with new resonant states in the top sector, consisting of a heavy top-philic scalar/pseudoscalar particle, a heavy $Z'$ boson, and a particularly challenging region for the top-squarks (the so called {\it top mass corridor}).

In Section \ref{sec:analysis} we assessed the discovery reach of a simple class of quantum observables based on spin correlations for resonant new physics. We have considered a simplified, yet realistic scenario for measurements in the near future. Our analysis shows that quantum observables, such as the entanglement markers $D$, can yield an advantage when searching for new physics over the "classical" observables commonly employed, in particular for cases where NP yields a significantly different spin state with respect to the SM background. We also show a new physics case, the vector resonance,  where quantum observables are not expected to improve over the already existing discovery power, because of the similarity with the SM in the respective spin states.

In summary, we find that in regions of parameter space where the invariant mass distribution of $t \bar t$ pair is very mildly affected by new physics, sizable effects in $t \bar t$ spin entanglement are present and can be detected. Our analysis further supports the idea that quantum observables provide an additional handle in the search for new phenomena and encourages more explorations in this direction. 

\section*{Acknowledgements}
\addtocounter{section}{1}

CS and ST thank Yoav Afik, Federica Fabbri, and James Howarth and FM thanks Luca Mantani and Rafael Aoude  for many enlightening discussions. 
CS and EV are supported by the European Union’s Horizon 2020 research and innovation programme under the EFT4NP project (grant agreement no.\@ 949451) and by a Royal Society University Research Fellowship through grant URF/ R1/ 201553. FM acknowledges support by FRS-FNRS (Belgian National Scientific Research Fund) IISN projects 4.4503.16. 
ST is supported by a FRIA (Fonds pour la formation \`a la Recherche dans l’Industrie et dans l’Agriculture) Grant of the FRS-FNRS (Belgian National Scientific Research Fund).

\clearpage

\appendix

\section{Calculation of the spin correlation matrix} \label{app:spin_calc}

The spin correlation matrix $\mathcal C$ for a given model can be calculated in a variety of ways. In this work, the $\mathcal C$'s are extracted from helicity amplitudes, using the method described in this Section. 

Helicity amplitudes  are obtained with  the insertion of suitable projectors in the spinor chain. For a massive particle, the projector for spin in direction $\hat{s}$ and sign $\sigma = \pm 1$ takes the form:
\begin{equation}
    P(\vec{s}, \sigma) = \frac{1 + \sigma \, \gamma^5 \, s^\mu \gamma_\mu }{2}. \label{spinproj}
\end{equation}
The four-vector $s^\mu$ is constructed from $\hat{s}$ such that:
\begin{equation}
    \begin{dcases}
    \vec{s} \, / \, |\vec{s}| = \hat s, \\
    s^\mu s_\mu = -1, \\
    s^\mu p_\mu = 0,
    \end{dcases} \label{s_conditions}
\end{equation}
where $p$ is the momentum of the particle. The system of equations \eqref{s_conditions} admits a unique solution, which defines the spin 4-vector $s^\mu$ up to the sign $\sigma$. Conventionally, the spin vector of an antiparticle is defined with a relative minus sign with respect to the spin vector of the corresponding particle, to comply with the usual requirement that for a (massless) particle helicity and chirality are identical, while for an antiparticle helicity and chirality are opposite.

The replacement:
\begin{equation}
    \psi \to P(\vec{s}, \sigma) \, \psi
\end{equation}
in all spinor chains, followed by the usual sum over helicities of $\psi$, then produces the scattering amplitude $\mathcal M(\vec{s}, \sigma) $ where the particle $\psi$ is spinning along direction $\vec{s}$ with sign $\sigma$.

The top and anti-top spin vectors enter the squared amplitude through dot products with other spin vectors and momenta and in contraction with $\varepsilon$ tensors, e.g.\@ $(s_{t})^\mu (p_{1})_\mu$ and $\varepsilon_{\mu \nu \rho \kappa} (p_{1})^\mu (p_{2})^\nu (s_{t})^\rho (s_{\bar{t}})^\kappa$. These can be readily evaluated in terms of $t \bar t$ kinematical quantities, once an explicit parameterisation of the $2 \to 2$ scattering has been chosen. For instance, using the helicity basis $\lbrace \hat k, \hat r, \hat n \rbrace$ in the $t \bar t$ frame one obtains for the momenta:
\begin{equation}
    \begin{dcases}
        p_1 = \frac{m_{t \bar t}}{2} \, ( 1, \, c_\theta, \, s_\theta, \, 0 ), \\
        p_2 = \frac{m_{t \bar t}}{2} \, ( 1, \, -c_\theta, \, -s_\theta, \, 0 ), \\
        p_t = \frac{m_{t \bar t}}{2} \,  ( 1, \, \beta, \, 0, \, 0), \\
        p_{\bar t} = \frac{m_{t \bar t}}{2} \, ( 1, \, -\beta, \, 0, \, 0 ), \\
    \end{dcases}
\end{equation}
and for the spin vectors:
\begin{equation}
    \begin{dcases}
        s_t(\hat{s}) = \sigma \, \frac{m_{t \bar t}}{2 m_t} \,  ( \beta \, \delta_{\hat{s} \hat{k}}, \, \delta_{\hat{s} \hat{k}}, \, \frac{2 m_t}{m_{t \bar t}} \, \delta_{\hat{s} \hat{r}}, \, \frac{2 m_t}{m_{t \bar t}} \delta_{\hat{s} \hat{n}} ),  \\
        s_{\bar t}(\hat{s}) =  \bar \sigma \, \frac{m_{t \bar t}}{2 m_t} \, ( -\beta \, \delta_{\hat{s} \hat{k}}, \, \delta_{\hat{s} \hat{k}}, \, \frac{2 m_t}{m_{t \bar t}} \delta_{\hat{s} \hat{r}}, \, \frac{2 m_t}{m_{t \bar t}} \delta_{\hat{s} \hat{n}} ). \\
    \end{dcases}
\end{equation}

The amplitude squared:
\begin{equation}
    \big| \, \mathcal M \big( t = (\vec i, \sigma); \ \bar t = (\vec j, \bar \sigma) \, \big) \big|^2
\end{equation}
is proportional to the probability to produce the top quark spinning along $\vec{i}$ with sign $\sigma$ and simultaneously the anti-top spinning along $\vec{j}$ with sign $\bar \sigma$. Spin correlation coefficients are then, by definition, given by:
\begin{align}
    \widetilde{C}_{ij} &= \sum_{\sigma = \pm 1} \sum_{\bar \sigma = \pm 1} \, \sigma \, \bar \sigma \, \big| \, \mathcal M \big( t = (\vec i, \sigma); \ \bar t = (\vec j, \bar \sigma) \, \big) \big|^2,\\
    A &= \sum_{\sigma = \pm 1} \sum_{\bar \sigma = \pm 1} \big| \mathcal M \big( t = (\vec i, \sigma); \ \bar t = (\vec j, \bar \sigma) \, \big) \big|^2.
\end{align}

\section{Generation details for SUSY in the top mass corridor} \label{app:susy_gen}

Simulating top-squark production in the top mass corridor scenario is notoriously challenging \cite{1804.00111}. We will assume that the decay into $\tilde{\chi}^0_1 \, b \, W$ is the dominant channel available for stops, so that the top squark is likely to be significantly narrower than its SM counterpart:
\begin{equation}
    \Gamma_{\tilde{t}_1} \ll \Gamma_{t}.
    \label{nwa_stops}
\end{equation}

The NWA for the top squark is valid across a wide range of coupling strengths and other supersymmetric parameters, and the process \eqref{susy_process} is effectively modelled as a sequence of on-shell productions and decays:
\begin{align}
     p \, p &\to \tilde{t}_1 \, {\tilde{t}}^*_1 \\
    &\tilde{t}_1 \to W^+ \, b \, \tilde{\chi}^0_1 , \ \ {\tilde{t}}^*_1 \to W^- \, \bar b \, \tilde{\chi}^0_1, \\
    &\hspace{3mm} W^+ \to \ell^+ \, \nu_\ell, \ \ W^- \to \ell^- \, \bar{\nu}_\ell.
\end{align}

If instead the SM top quark was the narrower particle, $\Gamma_{\tilde{t}_1} \gg \Gamma_{t}$, the appropriate decay chain would have been:
\begin{align}
     p \, p &\to t \, \bar{t} \, \tilde{\chi}^0_1 \, \tilde{\chi}^0_1\\
    &t \to W^+ \, b, \ \ \bar{t} \to W^- \, \bar b, \\
    &\hspace{3mm} W^+ \to \ell^+ \, \nu_\ell, \ \ W^- \to \ell^- \, \bar{\nu}_\ell,
\end{align}
while for $\Gamma_{\tilde{t}_1} \sim \Gamma_{t}$ the process $p \, p \to W^+ \, b \, W^- \, \bar b \, \tilde{\chi}^0_1 \, \tilde{\chi}^0_1$ should be considered as a whole, with subsequent decays of $W$ into leptons.  In all cases, using the NWA will only affect the process kinematics, the spin state is always evaluated exactly. \smallskip

\begin{figure}[t]
    \centering
    \includegraphics[width=0.48\textwidth]{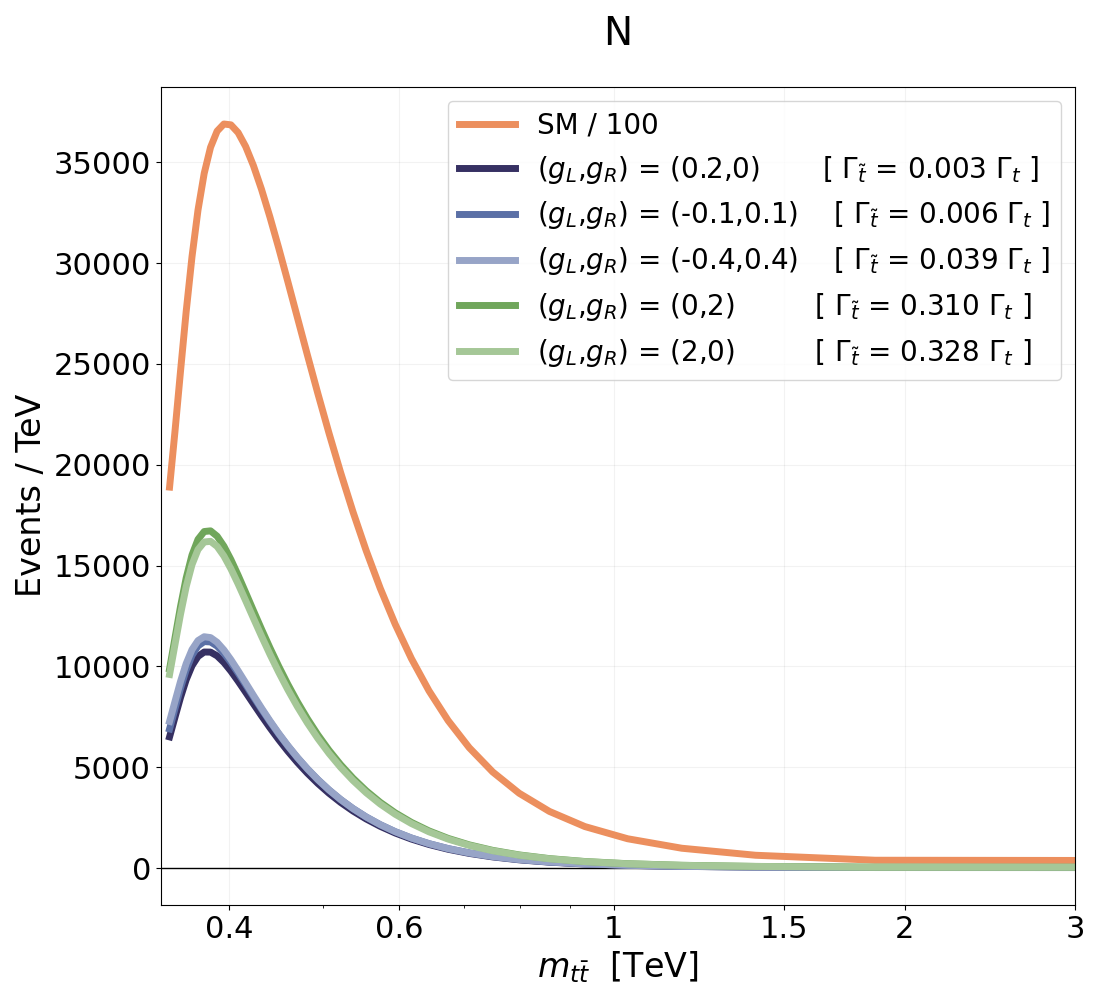}
    \includegraphics[width=0.48\textwidth]{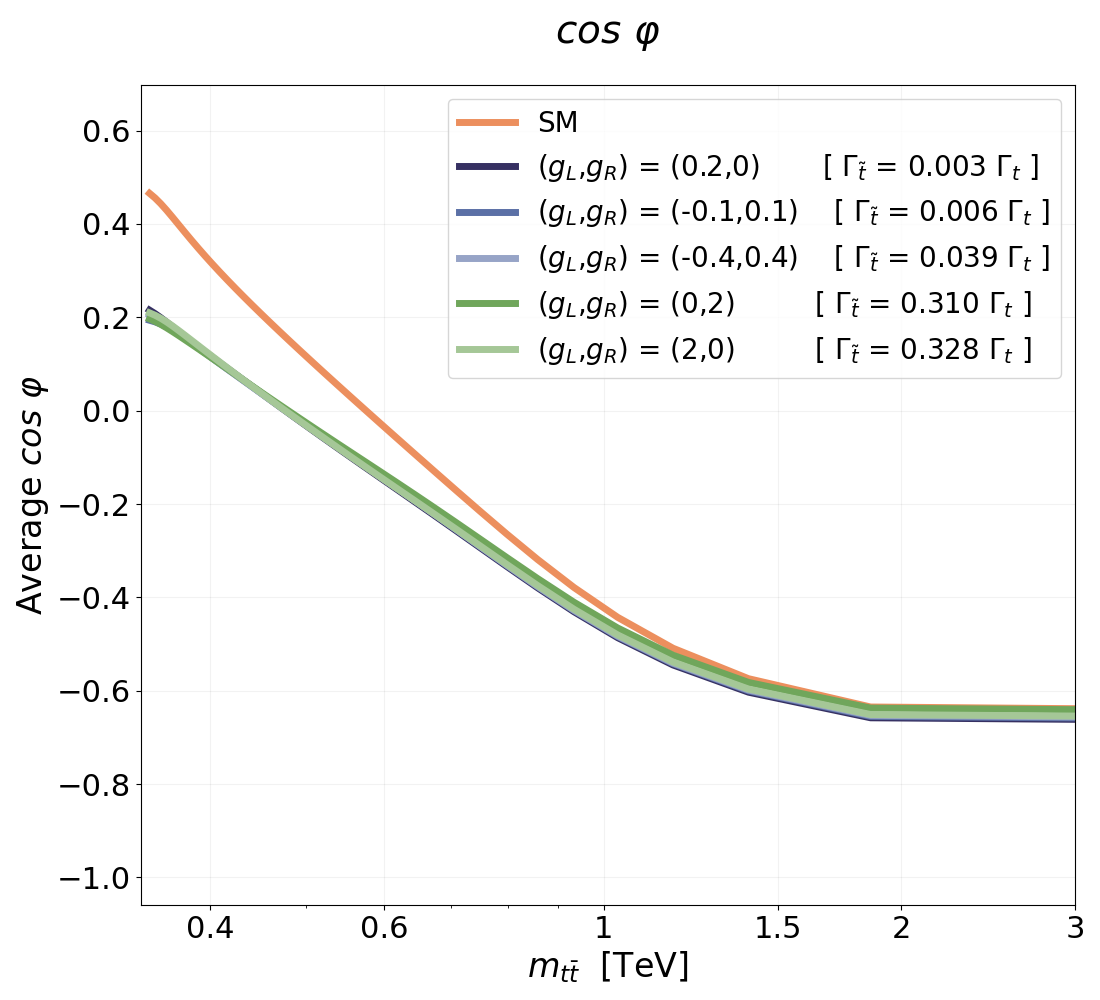}
    
    \vspace{5mm}
    \includegraphics[width=0.49\textwidth]{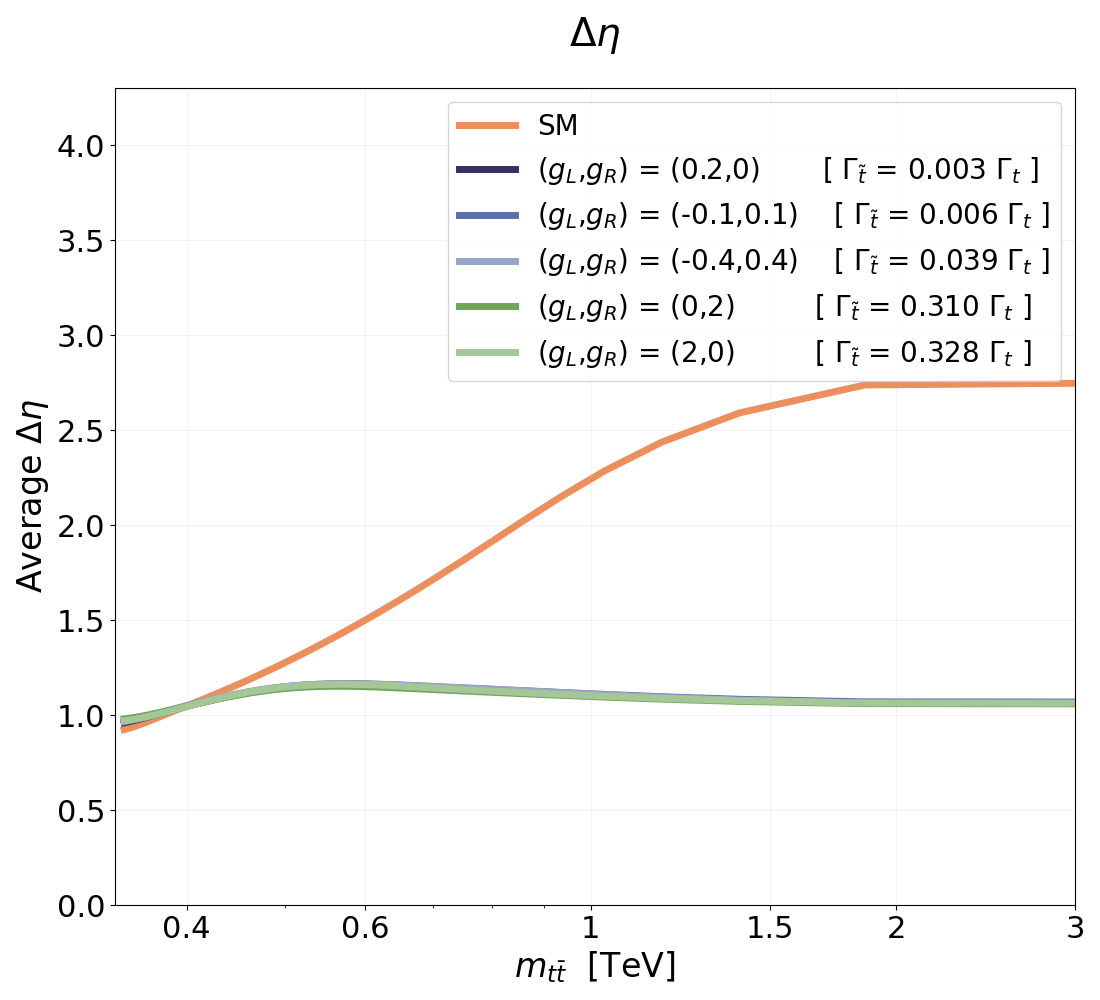}
    \includegraphics[width=0.49\textwidth]{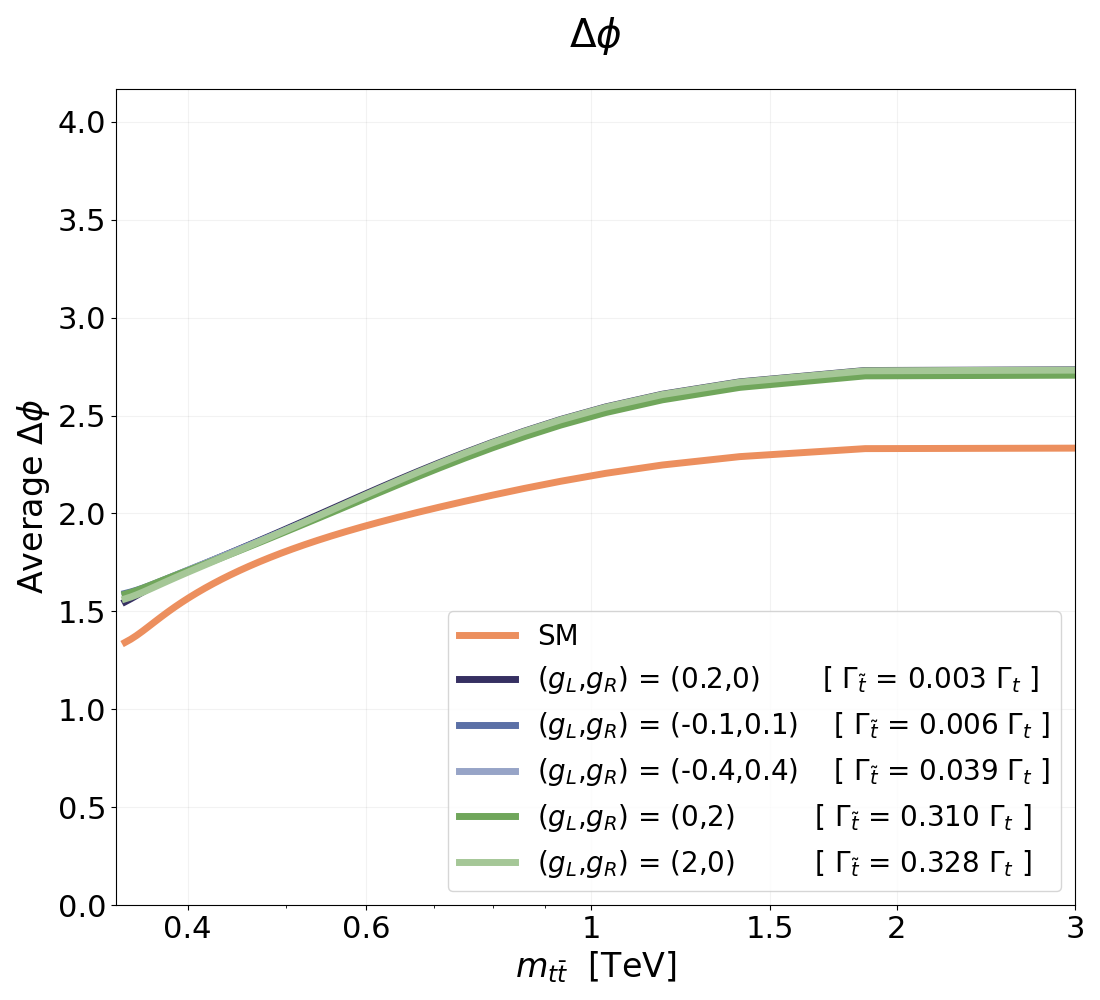}
    \caption{Observables considered in our analysis differential in the invariant mass of the $t \bar t$ system $m_{t \bar t} \equiv m_{b \bar b \ell \ell \nu \nu}$ in the SM and in SUSY top-squark pair production events with $m_{\tilde{t}} = 400 \, \text{GeV}$ and $m_{\chi_0^1} = 228 \, \text{GeV}$, for a variety of couplings $(g_L, g_R)$. Couplings for which the NWA is valid (blue) yield overlapping distributions, while those for which the NWA is not valid (green) yield overlapping distribution for spin observables but different cross-sections. Top row: number of events and $\cos \varphi$. Bottom row: $\Delta \eta$ and $\Delta \phi$. The entanglement markers $D$ are not plotted as they are consistent with zero in all SUSY cases we considered.}
    \label{fig:stop_nwa}
\end{figure}

In Figure \ref{fig:stop_nwa} we show explicitly that all observables we consider in our study are independent of the top-squark decay couplings in the NWA. In fact, spin observables, defined as a ratio of cross-sections, are largely independent of top-squark couplings even when the NWA is not valid.

In the NWA top-squark pair production and decay become independent of each other, and the rate for top-squark production is only a function of its mass (and of the QCD couplings, that in SUSY are the same as in the SM). Results for top-squark pair production in this limit are available at NLO+NNLL accuracy for a variety of collider scenarios \cite{Beenakker:2016lwe}.

\clearpage

\bibliographystyle{JHEP}
\bibliography{refs.bib}
\end{document}